\documentclass[11pt,a4paper]{article}
\usepackage[utf8]{inputenc}
\usepackage[T1]{fontenc}
\usepackage{lmodern}
\usepackage[margin=1in]{geometry}
\usepackage{amsmath,amssymb,amsfonts}
\usepackage{graphicx}
\usepackage{tikz}
\usepackage{caption}
\usepackage{subcaption}
\usepackage[hidelinks]{hyperref}
\usepackage[numbers,sort&compress]{natbib}

\title{Analysis and Simulation of Cooperative Buffer Strategies\\for Traffic Wave Mitigation}
\author{Raphael Korbmacher\thanks{University of Wuppertal, 42119 Wuppertal, Germany. E-mail: \href{mailto:korbmacher@uni-wuppertal.de}{korbmacher@uni-wuppertal.de}} \and
Daniel Straub\thanks{Technical University of Munich (TUM), 80333 Munich, Germany.} \and
Antoine Tordeux\footnotemark[1] \and
Claudia Totzeck\footnotemark[1]}
\date{}

\begin{document}
\maketitle
\begin{abstract}

Stop-and-go waves in traffic flow pose a persistent challenge, compromising safety, efficiency, and environmental sustainability. 
This paper presents a novel mitigation strategy identified through an exploratory multi-agent deep reinforcement learning experiment in a ring-road environment. 
In this experiment, the agents are observed to develop a driving policy in which most vehicles maintain minimal headways to maximize throughput, while a single "buffer" vehicle adopts a larger headway to absorb perturbations and prevent wave propagation. 
We demonstrate within a parametrised analytical framework, under a linearized long-wavelength string-stability criterion, that this strategy optimises overall flow among non-uniform time-gap distributions.
We further show through simulation that adapting this cooperative strategy to classical nonlinear car-following models, such as the Intelligent Driver and Full Velocity Difference models, yields improvements in stability, traffic efficiency, and speed oscillation across broad parameter sweeps. 
We additionally show that sharing the buffering role among multiple vehicles reduces the per-vehicle buffering time and mitigates intra-platoon instabilities in larger platoons, which may offer a more practically feasible strategy.

\end{abstract}


\section{Introduction}
\label{sec:1}

 Traffic jams represent an undesirable traffic condition with numerous negative consequences, including economic losses, increased fuel consumption, and higher pollutant emissions~\cite{nishi2013theory,li2014stop}. Causes of traffic jams can include bottlenecks, accidents, or construction sites, but can also arise for no apparent reason. These spontaneous jams are often referred to as phantom jams or, in this context, stop-and-go waves (SGW). Despite more than seven decades of research into stop-and-go dynamics, the emergence of self-sustaining waves in traffic flow remains incompletely understood.

The focus of this paper is on identifying strategies to prevent or mitigate the emergence of SGWs. Beyond strategic planning measures such as variable speed limits~\cite{hegyi2005optimal,hegyi2010dynamic}, the prevention of stop-and-go traffic has been approached from two main perspectives. The first is a microscopic one, which examines the behaviour of individual vehicles to determine if it can reduce or eliminate stop-and-go waves. These approaches are known as Jam-Absorption Driving (JAD) strategies. The second perspective is collective, termed traffic string stability analysis, where traffic is considered unstable if perturbations amplify along the flow. %upstream.

Although there are anecdotal reports of individuals mitigating stop-and-go waves through their driving behaviour~\cite{SGW_Blog}, JAD was not scientifically investigated until 2013 by Nishi et al.~\cite{nishi2013theory}. Two common JAD strategies that can foster SGW suppression are the "slow-in" and "fast-out" approach, as well as maintaining a large time gap between subsequent vehicles to avoid deceleration until the stop phase of the SGW has passed~\cite{wang2021jam}. Ideally, this allows the vehicle to hold and guide all following vehicles to approach an upcoming wave in a controlled manner, thereby achieving SGW suppression~\cite{he2025review}. However, these strategies are controversial and give rise to a \emph{first paradox}: while maintaining a high time gap may reduce SGWs in specific instances, the enlarged gap itself consumes road space and thereby lowers the capacity. This loss grows with the number of vehicles adopting the strategy.

Following years of theoretical investigations on the stability of traffic using car-following models \cite{orosz2010traffic,gasser2004bifurcation,orosz2004global}, the analysis of stability in real traffic situations with human drivers began with Sugiyama et al.~(2008)~\cite{sugiyama2008traffic} and their famous circuit road experiment. Their study demonstrated how a uniform flow of 22 vehicles on a closed-loop track could spontaneously transition into a traffic jam due to small fluctuations in driver behaviour, even without bottlenecks or external disruptions. Similar experiments reaching the same conclusions have been conducted by~\cite{tadaki2013phase,jiang2014traffic,stern2018dissipation}. These studies show that human driving behaviour is inherently unstable, leading to stop-and-go traffic without obvious causes. Consequently, expectations that technological solutions such as Adaptive Cruise Control (ACC) and autonomous vehicles can improve traffic stability rose \cite{treiber2001microsimulations,davis2004effect,bose2001analysis,korbmacher2025understanding}. While initial expectations were high, and experiments like Stern et al.~(2018)~\cite{stern2018dissipation} showed positive effects of autonomous vehicles on traffic stability, subsequent works have revealed that current market technologies are even more unstable than human drivers~\cite{makridis2018estimating,gunter2020commercially,makridis2020empirical,ciuffo2021requiem}. This leads to the \emph{second paradox}: 
achieving string-stable behaviour often comes at the cost of driving comfort, for instance through abrupt acceleration changes, leading manufacturers to prioritize comfort over stability (individual incentives over collective ones)~\cite{korbmacher2025understanding, gunter2020commercially,ciuffo2021requiem}.
%manufacturers must balance stable behaviour against comfort, often prioritizing the latter (individual incentives over collective ones)~\cite{korbmacher2025understanding, gunter2020commercially,ciuffo2021requiem}. 
Although technologically not yet realized, researchers agree that vehicle-to-vehicle communication will have a positive impact on mitigating SGW~\cite{de2004design,bu2010design,milanes2013cooperative}. 
This relates to so-called Cooperative ACC (CACC) systems, in which vehicles are connected and communicate with each other. 
These communications can significantly reduce the system response times, by eliminating the need for external sensors, which are prone to delays \cite{brunner2022comparing}. 
They can also enable vehicles to cooperate, for example by synchronising emergency braking or facilitating manoeuvres such as cutting in and out \cite{arnaout2011towards,kapsalis2025cooperative}.
Inter-vehicle cooperation could also mitigate SGW formation by assigning specific roles to each vehicle.

\begin{figure}[htbp]
    \centering\medskip%\medskip
    \begin{minipage}{0.48\textwidth}
        \centering
        \begin{subfigure}[t]{\linewidth}\centering \includegraphics[width=\linewidth, height=5cm]{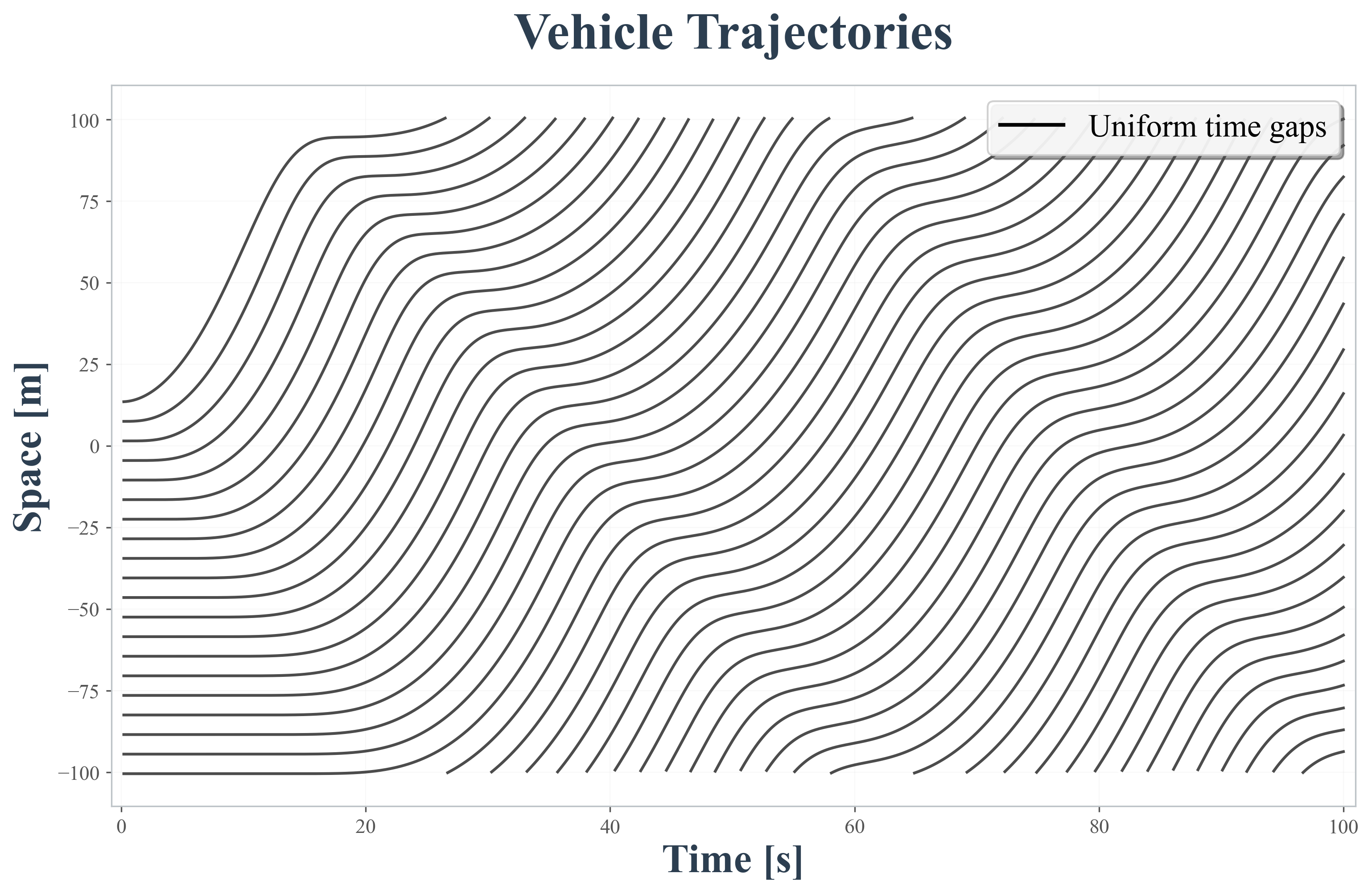}\caption{}\end{subfigure} \\[2.5mm]
        \begin{subfigure}[t]{\linewidth}\centering \includegraphics[width=\linewidth, height=5cm]{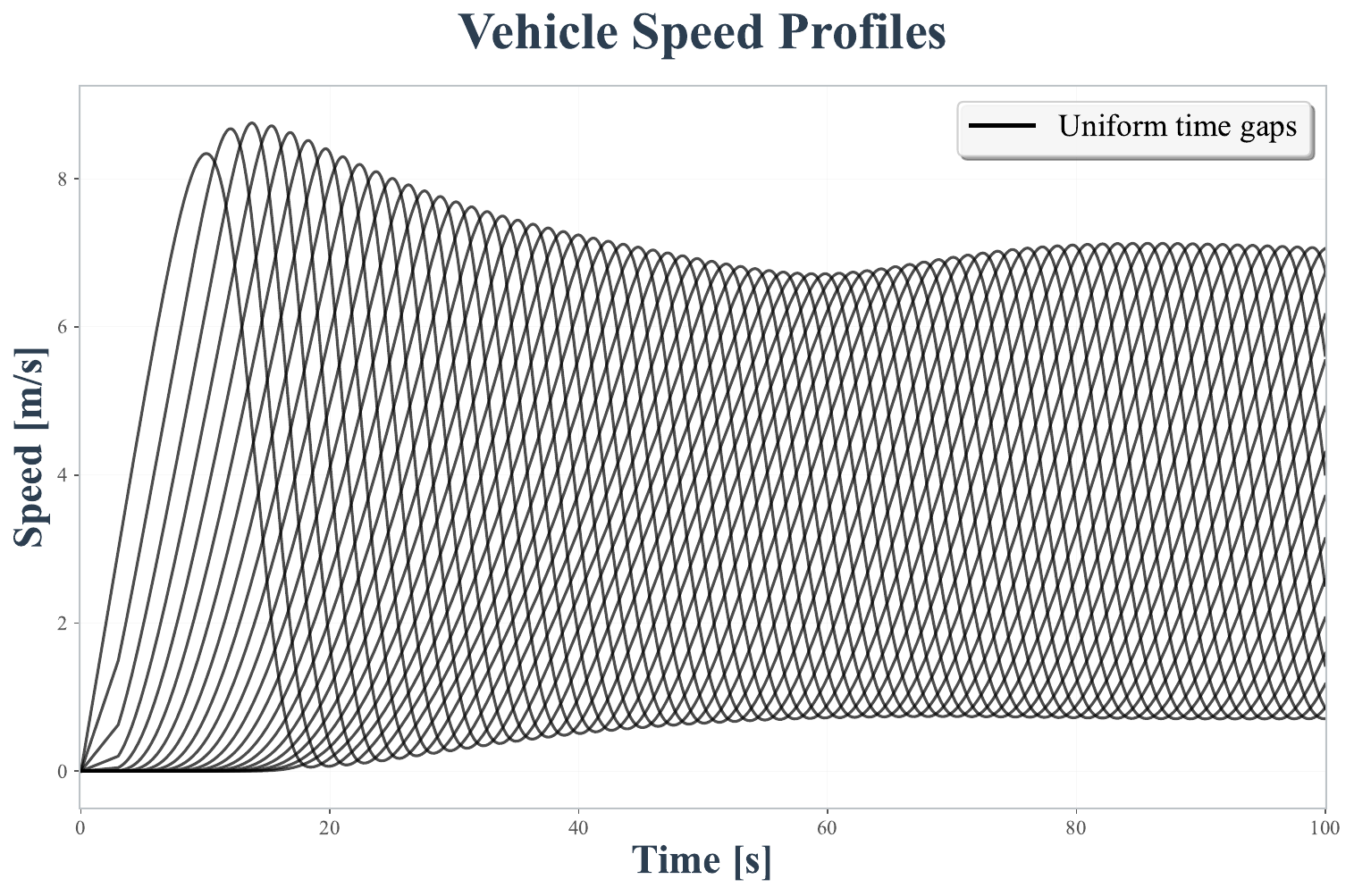}\caption{}\end{subfigure}
    \end{minipage}
    \hfill
    \begin{minipage}{0.48\textwidth}
        \centering
        \begin{subfigure}[t]{\linewidth}\centering \includegraphics[width=\linewidth, height=5cm]{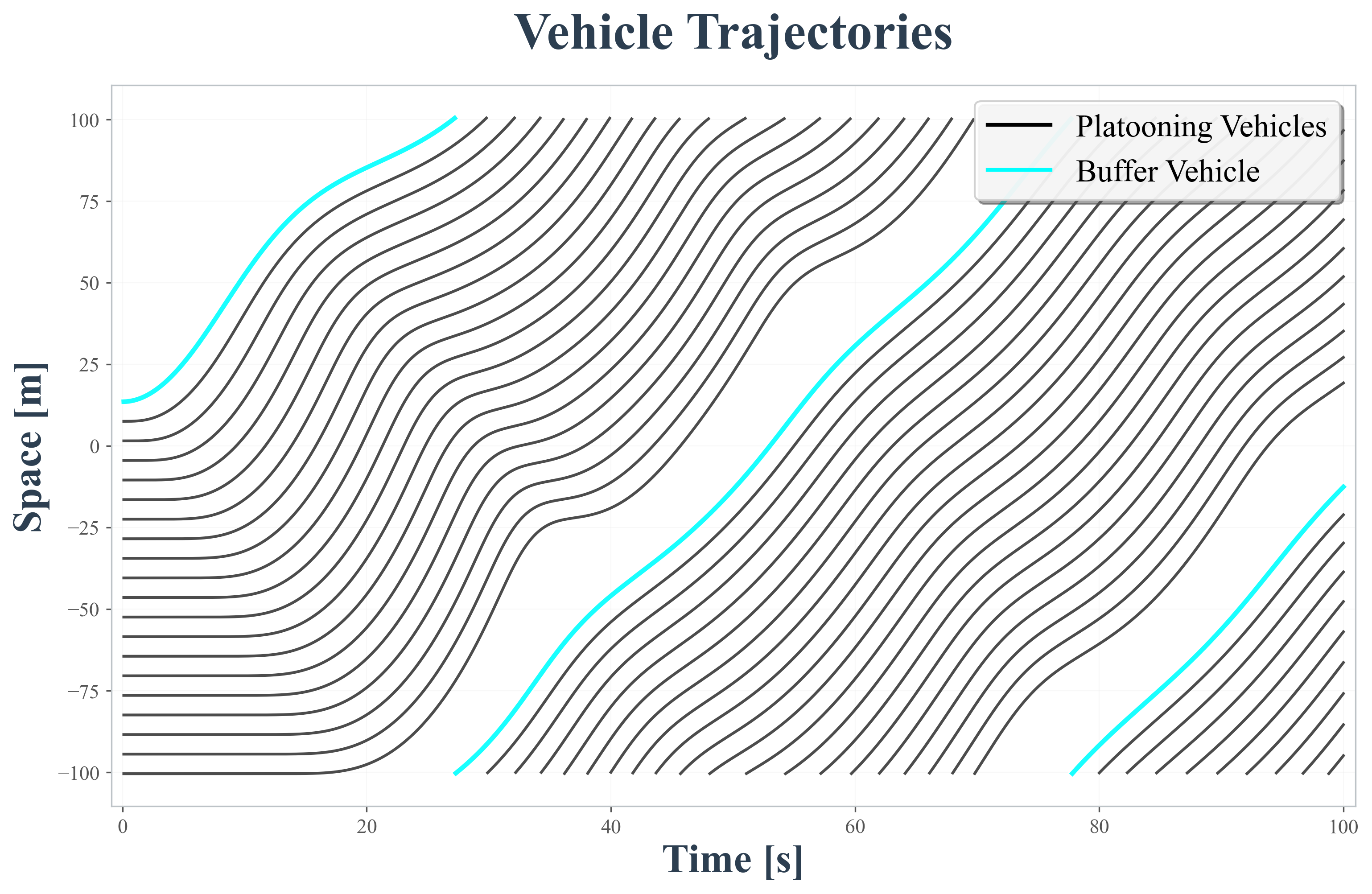}\caption{}\end{subfigure} \\[5mm]
        \begin{subfigure}[t]{\linewidth}\centering \includegraphics[width=\linewidth, height=5cm]{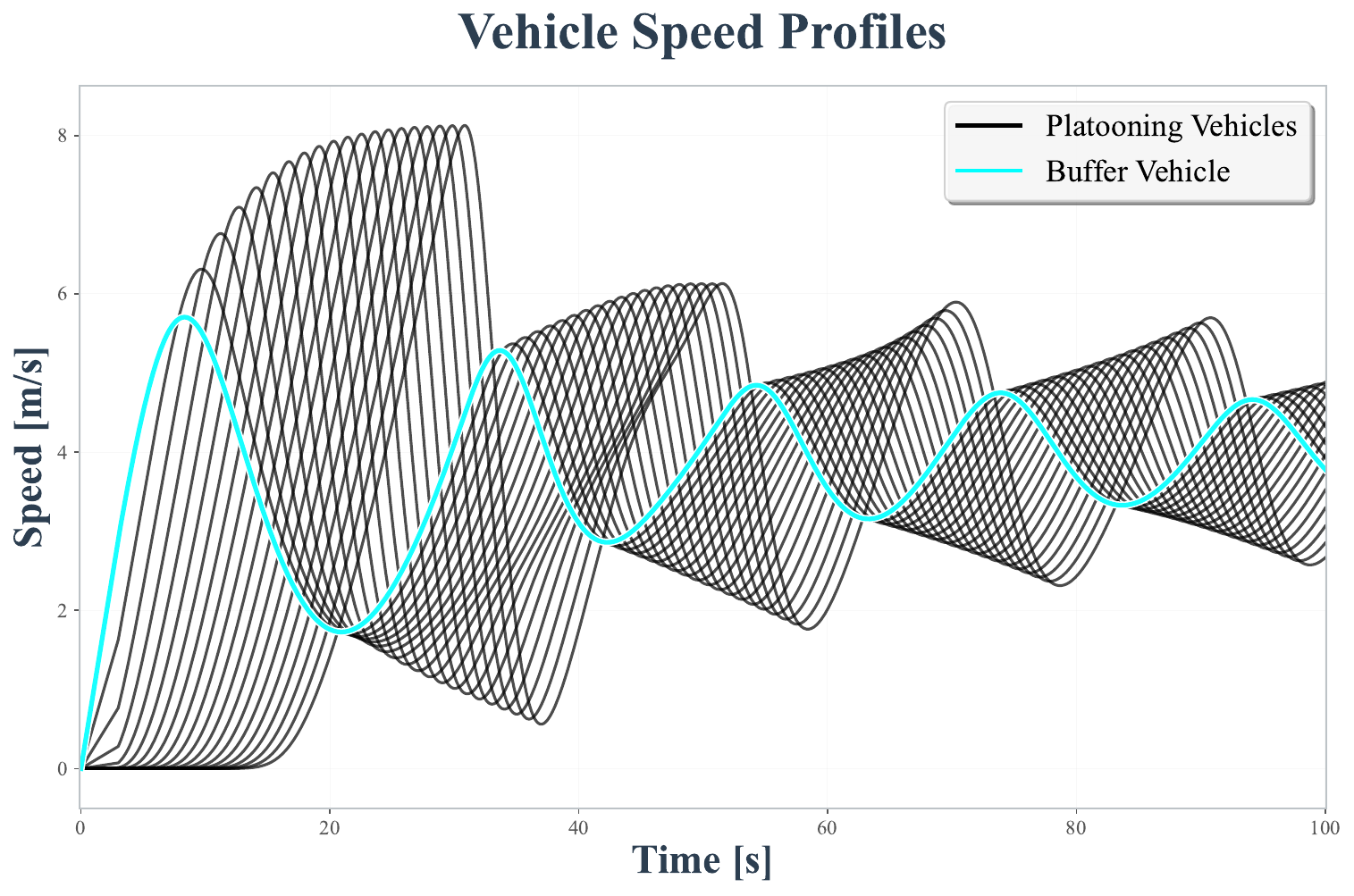}\caption{}\end{subfigure}
    \end{minipage}%\medskip
    \caption{Simulation results using identical IDM parameter settings, without (left panels) and with (right panels) the cooperative buffering strategy. Both scenarios start from the same initial traffic jam condition. In Fig.~\ref{fig_strategy}(a), the trajectories of 20 vehicles with uniform time gaps are shown, with the corresponding speed profiles in Fig.~\ref{fig_strategy}(b). A stop-and-go wave emerges and propagates upstream. The right panels display the same setup but with the cooperative buffering strategy applied. In this case, the system converges to a laminar flow without wave formation. The trajectories can be computed and visualized online; see~\cite{SimBufferingStrategy}.}
    \label{fig_strategy}
\end{figure}

In this contribution, we train Deep Reinforcement Learning (DRL) agents in an experimental setup similar to Sugiyama's experiment. 
Through this exploratory experiment a buffering strategy that helps mitigate both paradoxes is identified. 
While the strategy imposes a stability condition on the time gaps, it retains a degree of freedom in the calibration of comfort-related parameters, such as acceleration and deceleration bounds, particularly when the sum of squared time gaps is large.
Rather than homogeneous traffic, where all vehicles behave in the same way, the DRL agents exhibit heterogeneous/non-uniform behaviour. One vehicle acts as a \emph{buffer}, maintaining a large time gap, while the others follow closely behind, as in a \emph{platoon} (see the proof-of-concept in Figure~\ref{fig_strategy}). 
We present in Section~\ref{sec:3} a parametrised framework in which we demonstrate that this strategy optimises overall flow under linear, long-wavelength string stability constraints. We then adapt in Section~\ref{sec:4}  the cooperative buffer/platoon strategy to classical nonlinear car-following models --- namely the Intelligent Driver Model and the Full Velocity Difference Model --- and evaluate its performance through simulation, showing improvements in both stability and traffic efficiency. We conclude in Section~\ref{sec:5} by discussing the limitations and implications of these findings for the development of future autonomous vehicle systems.

\section{Discovery of Cooperative Buffering Strategy via Multi-Agent DRL\label{sec:2}}

Classical car-following theory generally assumes that all vehicles follow identical driving behaviour. This assumption underlies the vast majority of traffic regulation models and, through symmetry, restricts the space of admissible stability-optimal solutions to uniform configurations. The present section motivates the removal of this assumption. However, allowing each vehicle to adopt an individually tailored driving policy increases the dimensionality of the joint policy space proportionally to the number of vehicles, rendering classical optimisation approaches intractable. Multi-agent deep reinforcement learning (DRL) is particularly well suited to address this challenge: rather than exhaustively searching a high-dimensional parameter space, it allows agents to discover effective policies through direct interaction with the environment, scaling naturally to large numbers of vehicles. We therefore conduct an exploratory experiment in which the space of heterogeneous driving policies is searched by multi-agent DRL without imposing homogeneity a priori, with the aim of identifying cooperative behaviours that would be difficult to find by conventional means. The DRL experiment serves exclusively as an exploratory discovery tool that motivates the analytical and numerical investigations of Sections III and IV.

\subsection{Problem Formulation}
The car-following task is formulated as an $N$-agent Markov decision process~\cite{franccois2018introduction} on a closed ring road of length $L_{\mathrm{DRL}} = 220$\,m populated with 15 to 25 vehicles of effective length $\ell = 5$\,m, with integration time step $\delta t = 0.1$\,s. The ring topology follows the Sugiyama set-up~\cite{sugiyama2008traffic} and supports the spontaneous emergence of stop-and-go oscillations under periodic boundary conditions. The environment is implemented within the Gymnasium framework~\cite{towers2024gymnasium}.

All agents use the same reward function, observation and action spaces, network architecture and training conditions, but each agent has its own set of policy parameters. The agents are trained sequentially with a single-agent DRL algorithm.
At each time step, %$t$ with simulation timestep $\delta t = 0.1$ s, 
the DRL agent observes a four-dimensional state vector, containing its own velocity, % $v_t$, 
the relative distance to the front vehicle, %$d_t$, 
the relative velocity to the front vehicle, %$\Delta v_t$, 
and the front vehicle's velocity. %$v_{t,f}$. 
Including both the relative velocity and the velocity of the predecessor %$v_{t,f}$ 
provides redundant but potentially stabilizing information to the neural network.
The action space is continuous, consisting of a single acceleration command $a_t \in [-5, 2]$ m/s$^2$, as per ISO 15622 standards for adaptive cruise control systems~\cite{ISO15622_2018}.
For policy approximation, a Proximal Policy Optimization (PPO)~\cite{schulman2017proximal} is utilized, selected for its sample efficiency and stability in continuous control tasks. The policy network is a multi-layer perceptron (MLP) with three hidden layers of 64 units each, using ReLU activations. Hyperparameters include a learning rate of $3 \times 10^{-4}$, discount factor $\gamma = 0.99$, GAE lambda of 0.95, clip range of 0.2, entropy coefficient of 0.01, and value function coefficient of 0.5, tuned empirically for convergence. More information is available at \url{https://github.com/RaphaelVZU/DRL_Vehicle_SGW_Experiement}.

\subsection{Training}
A crucial aspect of the DRL approach is the design of a reward function that balances multiple objectives: safety, efficiency, comfort, and stability. In the experiments, for each vehicle in the DRL model, the reward $r_t$ at time $t$ is defined as:
\begin{equation}
r(t) = w_v r_v(t) + w_d r_d(t) + r_s(t),
\end{equation}
where the velocity reward $r_v(t) = -|v(t) - v_0|/v_d$ penalizes deviation from the desired velocity $v_0 = 10$ m/s and is assigned weight $w_v=5$. The distance reward $r_d(t) = -|d(t) - d^*(t)|/d^*(t)$ encourages maintaining the desired headway $d^*(t) = T v(t) + \ell$, with $T= 1$~s the time gap, an assigned weight $w_d=7.5$, and $\ell>0$ the minimum spacing. The safety reward $r_s(t)$ is a piecewise-constant term taking small positive values (on the order of $+1$ to $+2$) in safe spacing regimes, zero in a neutral band, and large negative values (down to $-50$) as the gap approaches collision.

To enhance robustness, perturbations during training, random emergency braking events applied to random vehicles with probability 0.03 per step, lasting up to 30 steps (3 s) at 80\% of maximum deceleration ($-4$ m/s$^2$) are introduced. In addition, Gaussian noise ($\mathcal{N}(0, 0.5)$ m) is added to distance measurements to simulate sensor inaccuracies.
Each agent was trained for 200,000 time steps, using a batch size of 64~\cite{DRL_code}.

\subsection{Observed Cooperative Buffering Strategy}
After training, the learned behavior is evaluated in deterministic roll-outs on the ring by observing the achieved mean speed, the mitigation of SGWs, and the robustness against perturbations.
Notably, in the exploratory experiment, the configuration that achieved the highest mean speed while retaining robustness to perturbations is observed to be non-uniform. In this final configuration, one vehicle settles at a substantially enlarged time gap while the remaining $N-1$ vehicles operate in a tightly compressed platoon. The vehicle adopting the enlarged gap will henceforth be referred to as the \emph{buffer vehicle}. Since all agents are trained with an identical reward function and under identical conditions, this behavioural differentiation is not prescribed by the set-up, but arises during the training. At the operational level, the observed pattern is an asymmetric time-gap allocation. The configuration is collectively beneficial only if the buffer vehicle forgoes closing its gap, thereby stabilizing the entire traffic flow, while the remaining vehicles accept compressed time gaps to preserve global throughput. In fact, if other vehicles also increased their time gaps, this strategy would lose its efficiency, which is why modifying the behaviour of the buffer vehicle alone is insufficient. %which is why altering the behaviour of only the buffer vehicle is insufficient.
This empirical observation motivates the analytical and numerical investigation that follows: Section~\ref{sec:3} establishes why a non-uniform configuration is stability-optimal within a parametrised framework, while Section~\ref{sec:4} adapts and evaluates the strategy on classical nonlinear car-following models through simulation.
% This empirical observation motivates two questions that the remainder of the manuscript addresses analytically: why a non-uniform configuration is a stability-optimal (Section III) and how it should be parameterised (Sections IV–V).

\section{Mathematical Formalization\label{sec:3}}% of Optimal Cooperative Buffering\label{sec:3}}

A fundamental advantage of the DRL approach is that it does not impose the classical assumption of behavioural uniformity a priori. Although all agents are trained under identical rules, the observed configuration is non-uniform. Building on this observation, this section breaks with the classical assumption of uniform behaviour in car-following models. Using a parametrised linear framework, we mathematically prove that extreme non-uniform behaviour---specifically, buffering---constitutes the globally optimal solution that maximises system performance under linearised long-wavelength string stability constraints.

\subsection{The Optimal Stability Control Problem}

Consider $N\ge2$ vehicles navigating a one-dimensional segment of length $L>0$ subject to periodic boundary conditions. The longitudinal dynamics of the platoon can be linearly described by a general non-uniform car-following model. The acceleration of the $n$-th vehicle is governed by:
\begin{equation}
\frac{d}{dt}v_n(t) = a_n d_n(t) + b_n v_n(t) + c_n v_{n+1}(t),
\label{eq:linear_model}
\end{equation}
where $v_n(t) = \frac{d}{dt}x_n(t)$ is the velocity of the $n$-th vehicle, and $d_n(t) = x_{n+1}(t) - x_n(t)$ is the distance headway to the leading vehicle. The parameters satisfy $a_n > 0$, $b_n < 0$, $c_n\ge 0$, and
$b_n+c_n<0$ for all $n\in\{1,\dots,N\}$.

The uniform equilibrium solution corresponds to a stationary state where all vehicles move with an identical velocity $v$ and maintain constant spacings $d_n$, such that their respective accelerations are exactly zero. This requires:
\begin{equation}
a_n d_n + b_n v + c_n v = 0 \quad \forall n \in \{1, \dots, N\}.
\label{eq:Equilibrium}
\end{equation}

According to classical traffic flow theory, this uniform equilibrium solution is string stable in the long-wavelength limit if the system satisfies the following condition \cite{ngoduy2015effect}:
\begin{equation}
\sum_{n=1}^N \frac{b_n^2 - c_n^2}{2a_n^2} \ge \sum_{n=1}^N \frac{1}{a_n}.
\label{eq:general_stability}
\end{equation}
Note the condition simplifies to the well-established condition $b^2-c^2\ge2a$ for uniform parameters. 
For the scope of the analytical derivation, we now parametrize the equilibrium through the time gap $T_n:=g_n/v$ of each vehicle.
This corresponds to the \emph{constant time gap} (CTG) policy, a common principle in many car-following models, whereby the equilibrium gap is proportional to the speed. 
The policy is historically motivated by a rule attributed to the \emph{California Motor Vehicle Code} Pipes~\cite{pipes1953operational,Chandler1958}. 
The same principle ensures safe following behavior and remains in use in adaptive cruise control (ACC) systems. 
%Indeed, the ISO 15622 standard for ACC systems formalizes it, requiring a minimum selectable time gap
%$\tau_{\min}\ge 0.8$~s, with at least one setting offered in the range $\tau\in[1.5,2.2]$~s for higher speeds~\cite{ISO15622_2018}.
Using \eqref{eq:Equilibrium}, the time gap of vehicle $n$ is
\begin{equation}
\begin{aligned}
&T_n=\frac{g_n}{v}=-\frac{b_n+c_n}{a_n}\\
&\Longleftrightarrow\quad
b_n=-a_nT_n-c_n, \quad \forall n\in\{1,\dots,N\}.
\end{aligned}
\label{eq:CondPara}
\end{equation}
Hence, the CTG policy is not an additional assumption but a direct consequence of the linear model \eqref{eq:linear_model}. 
The time gap is entirely determined by the model parameters $(a_n,b_n,c_n)$ and, conversely, can be prescribed by choosing them.
Allowing $T_n$ to differ across vehicles accounts for heterogeneous driver or ACC settings.
Substituting the expression \eqref{eq:CondPara} into \eqref{eq:general_stability}, and using
$b_n^2-c_n^2=a_n^2T_n^2+2a_nc_nT_n$, the stability condition becomes
\begin{equation}
\sum_{n=1}^N \Big(T_n^2 + 2\,T_n\,\frac{c_n}{a_n}\Big) \ge \sum_{n=1}^N \frac{2}{a_n}=: \alpha_N > 0.
\label{eq:stability_constraint2}
\end{equation}
The threshold $\alpha_N$ is positive since $a_n>0$.

At equilibrium, all the gaps satisfy $g_n=vT_n$.
On the closed road, the total gap is conserved:
$\sum_{n=1}^N g_n = L-N \ell =: \bar L$, where $\bar L>0$ is the total free space.
The equilibrium speed is therefore determined by the time gaps:
\begin{equation}
\sum_{n=1}^N vT_n=\bar L
\quad\iff\quad
v(T_1,\dots,T_N)=\frac{\bar L}{\sum_{n=1}^N T_n}.
\label{eq:mean_speed}
\end{equation}

Since $v$ is inversely proportional to $\sum_n T_n$, maximizing the flow $q=Nv/L$ (for fixed $N$ and $L$) is equivalent to minimizing the total time gap $\sum_n T_n$. 
The stability condition \eqref{eq:stability_constraint2}, however, imposes a competing requirement: the sum
of squared time gaps must exceed a model-dependent threshold. This tension between an $L_1$ objective and an $L_2$ constraint makes the problem non-trivial and gives it its characteristic geometry. 

For a general heterogeneous fleet, let $\beta_n:=c_n/a_n\ge 0$ and recall that
$\alpha_N=\sum_{n=1}^N 2/a_n>0$. By \eqref{eq:stability_constraint2}, the optimal stability
control problem reads:

\smallskip
\textbf{Minimize:}
\begin{equation}
J(T_1,\dots,T_N)=\sum_{n=1}^N T_n
\label{eq:objective}
\end{equation}

\textbf{Subject to:}
\begin{align}
&\sum_{n=1}^N \big(T_n^2+2\beta_n T_n\big) \ge \alpha_N \quad \text{(stability constraint)}, \label{eq:constraint_1}\\[1mm]
&T_n \ge T_\text{min}, \quad \forall n\in\{1,\dots,N\}, \label{eq:constraint_2}
\end{align}
where $T_\text{min}>0$ is the minimum admissible time gap, assumed small enough that
\begin{equation}
\begin{aligned}
C&:=\alpha_N-\sum_{n=1}^N\big(T_\text{min}^2+2\beta_nT_\text{min}\big)\\
&=\alpha_N-NT_\text{min}^2-2T_\text{min}\sum_{n=1}^N\beta_n>0.
\end{aligned}
\label{eq:constraint_3}
\end{equation}
Assumption \eqref{eq:constraint_3} states that the configuration $T_n=T_\text{min}$ for all $n$
violates the stability constraint. Without it, this configuration would be admissible and
trivially optimal. For $c_n=0$ we have $\beta_n=0$, the stability constraint reduces to
$\sum_n T_n^2\ge\alpha_N$, and \eqref{eq:constraint_3} reads $\alpha_N>NT_\text{min}^2$.
The feasible set is the region $\{T_n\ge T_\text{min}\ \forall n\}$ lying outside the open ball of center $-(\beta_1,\dots,\beta_N)$ and radius $\sqrt{\alpha_N+\sum_n\beta_n^2}$. It is therefore non-convex, which is the source of the
structure of the solution derived below.

\subsection{The Non-Cooperative Optimal Solution}
Traditionally, traffic modelling assumes uniform behaviour. 
We refer to the uniform allocation, in which all vehicles hold the same time gap $T_n = T_\text{non}$, as the non-cooperative solution, since it requires no coordination among the vehicles.
To formalize this mathematically and evaluate its optimality, the Lagrangian of the optimization problem is set up:
\begin{equation}
\mathcal{L} = \sum_{n=1}^N \big(T_n + \mu_n(T_\text{min} - T_n)\big)
- \lambda \left(\sum_{n=1}^N (T_n^2+2\beta_nT_n) - \alpha_N\right),
\end{equation}
where $\beta_n:=c_n/a_n\ge0$, $\lambda \ge 0$ is the Lagrange multiplier for the stability constraint, and $\mu_n \ge 0$ are the Karush-Kuhn-Tucker (KKT) multipliers for the lower bound constraints. 
The first-order necessary optimality conditions yield:
\begin{equation}
\begin{aligned}
&\frac{\partial \mathcal{L}}{\partial T_n} = 1 - \mu_n - 2\lambda (T_n+\beta_n) = 0\\
&\implies T_n+\beta_n = \frac{1 - \mu_n}{2\lambda}.
\end{aligned}
\end{equation}
Under the assumption of a symmetric, uniform allocation, all vehicles are treated alike, leading to $\mu_n = \mu_m$ for all $n, m$.
We adopt $T_n\equiv T_\text{non}$ as the behavioural definition of the non-cooperative strategy in general, independently of possible heterogeneity in $\beta_n$.

Enforcing the active stability constraint (\ref{eq:constraint_1}) at the boundary to maximize efficiency, $NT^2+2T\sum_n\beta_n=\alpha_N$, yields the optimal uniform time gap:
\begin{equation}
\begin{aligned}
&T_n=T_\text{non}^* = -\bar\beta+\sqrt{\bar\beta^{\,2}+\frac{\alpha_N}{N}},\\
& \bar\beta:=\frac1N\sum_{n=1}^N\beta_n,
\quad\forall n\in\{1,\ldots,N\}.
\end{aligned}
\label{eq:T_non}
\end{equation}
which reduces to $T_\text{non}^*=\sqrt{\alpha_N/N}$ when $c_n=0$.
Substituting this back into (\ref{eq:mean_speed}), the corresponding optimally stable mean
speed for a uniform fleet is:
\begin{equation}
v_\text{non}^* = \frac{\bar{L}}{N\,T_\text{non}^*}
\qquad\Big(=\frac{\bar L}{\sqrt{N\alpha_N}}\ \text{ if } c_n=0\Big).
\label{eq:v_non}
\end{equation}
This represents the absolute limit of uniform traffic flow. 
Any uniform time gap smaller than $T_\text{non}^*$ violates condition (\ref{eq:constraint_1}) and transits into an unstable regime. 
Thus, while stable, the uniform solution represents merely a candidate constrained by the artificial requirement of uniform behaviour, whose optimality among all admissible (non-uniform) configurations is examined next.

\subsection{The Cooperative Optimal Solution}

We now challenge the foundational assumption of uniformity. To find the global optimum, non-uniform time-gap distributions must be explored. 
Write
\begin{equation}
T_n=T_\text{min}+u_n,\qquad u_n\ge 0,\qquad s:=\sum_{n=1}^N u_n .
\end{equation}
The objective becomes $J=NT_\text{min}+s$, so minimizing $J$ is equivalent to minimizing the
total excess gap $s$. Substituting into \eqref{eq:constraint_1} and using the definition
\eqref{eq:constraint_3} of $C$, the stability constraint becomes
\begin{equation}
\sum_{n=1}^N u_n^2+2\sum_{n=1}^N m_nu_n\ \ge\ C,
\quad m_n:=T_\text{min}+\beta_n .
\label{eq:u_constraint}
\end{equation}
Let $m^*:=\max_n m_n=T_\text{min}+\beta_{\max}$ with $\beta_{\max}:=\max_n\beta_n$. Since $u_n\ge0$,
\begin{equation}
\sum_{n=1}^N m_nu_n\le m^*s,
\quad
\sum_{n=1}^N u_n^2\le\Big(\sum_{n=1}^N u_n\Big)^2=s^2 .
\label{eq:inequality}
\end{equation}
The second inequality holds because the cross terms $u_nu_m$ are non-negative. Combining
\eqref{eq:u_constraint} and \eqref{eq:inequality}, every admissible configuration satisfies
\begin{equation}
s^2+2m^*s\ \ge\ C .
\end{equation}
The function $s\mapsto s^2+2m^*s$ is increasing on $s\ge0$ and vanishes at $s=0<C$. Hence
\begin{equation}
s\ \ge\ s^*:=-m^*+\sqrt{(m^*)^2+C}\ >0 .
\label{eq:s_star}
\end{equation}

The bound \eqref{eq:s_star} is attained if and only if both inequalities in
\eqref{eq:inequality} are equalities. The second one, $\sum_n u_n^2=(\sum_n u_n)^2$, holds for
non-negative variables if and only if at most one $u_n$ is non-zero. 
The first one holds if and only if every $n$ with $u_n>0$ satisfies $m_n=m^*$, i.e., $\beta_n=\beta_{\max}$. 
Under \eqref{eq:constraint_3}, the optimal stability control problem has a minimizer. 
The minimizers are exactly the configurations
\begin{align}
&~\;T_n^* = T_\text{min}, \qquad n\neq n_0, \label{eq:opt_platoon}\\
&\begin{array}{ll}
T_{n_0}^* \!\!\!&= T_\text{min}+s^*\\
&=-\beta_{\max}+\sqrt{\beta_{\max}^2+\alpha_N-\sum_{n\neq n_0}\big(T_\text{min}^2+2\beta_nT_\text{min}\big)},
\end{array}
\label{eq:opt_buffer}
\end{align}
where the ``buffer vehicle'' $n_0$ belongs to $\arg\max_n\, c_n/a_n$. For $c_n=0$, every
index is admissible as buffer vehicle, and \eqref{eq:opt_buffer} reads
$T_{n_0}^*=\sqrt{\alpha_N-(N-1)T_\text{min}^2}$.

Indeed, by the chain of inequalities above, every admissible configuration satisfies $s\ge s^*$. The
configuration \eqref{eq:opt_platoon}--\eqref{eq:opt_buffer} is admissible (it saturates
\eqref{eq:u_constraint}) and has $s=s^*$, so it is a global minimizer. Conversely, if $s=s^*$,
all inequalities in the chain are equalities, which forces the structure described above.
The expression in \eqref{eq:opt_buffer} follows from
$(T_\text{min}+\beta_{\max})^2+C=\beta_{\max}^2+\alpha_N-\sum_{n\ne n_0}(T_\text{min}^2+2\beta_nT_\text{min})$.

The maximum achievable stable throughput thus requires an extreme behavioral non-uniformity:
$N-1$ vehicles form a dense, high-capacity platoon operating at the minimum time gap
$T_\text{min}$, while a single cooperative vehicle adopts a large time gap and acts as a
buffer that stabilizes the whole fleet. When the vehicles are heterogeneous, the buffer role
is assigned to the vehicle with the largest ratio $c_n/a_n$, i.e., the one for which an
increase of the time gap yields the largest gain in stability. The corresponding mean speed is
\begin{equation}
v_\text{coop}^*=\frac{\bar L}{(N-1)T_\text{min}+T_{n_0}^*}.
\end{equation}

Then, under \eqref{eq:constraint_3}, $v_\text{coop}^*>v_\text{non}^*$.
Indeed, the uniform solution $T_n\equiv T_\text{non}^*$ is admissible, and therefore $\sum_n T_n^*\le N\,T_\text{non}^*$ by optimality of the cooperative
solution. Since $N\ge2$ and $T_\text{non}^*>T_\text{min}$, the uniform configuration has at least two components strictly above $T_\text{min}$, so it is not a minimizer. 
The inequality is therefore strict, and the result follows from \eqref{eq:mean_speed}.

\section{Nonlinear Microscopic Validation Framework\label{sec:4}}

% While the analytical proof in Section III guarantees the global optimality of a non-uniform time-gap distribution under linearized assumptions, real-world traffic is governed by complex, nonlinear dynamics. 
% In addition, the string stability condition \eqref{eq:general_stability} only holds for the longest wavelength. However, shorter wavelength modes can become unstable for non-uniform models \cite{treiber2012traffic}.
% To demonstrate that the cooperative buffering strategy effectively can dissipate finite-amplitude perturbations and expands the basin of attraction for laminar flow, the system is evaluated using two distinct second-order car-following models. %By demonstrating the strategy's efficacy across these different models, we provide evidence that the benefits of non-uniformity are not an artefact of a particular model.

While the analytical results in Section~\ref{sec:3} establish the global optimality of a non-uniform time-gap distribution under linearized, long-wavelength assumptions, real-world traffic dynamics are nonlinear and finite-amplitude perturbations are common. 
Moreover, the string stability condition \eqref{eq:general_stability} only guarantees stability for the longest wavelength mode. 
Shorter-wavelength modes can nonetheless become unstable for non-uniform models \cite{treiber2012traffic}. 
These theoretical results should therefore be regarded as an analytical benchmark rather than a directly deployable guarantee. 
To assess whether the cooperative buffering strategy remains effective for all wavelengths and beyond the linear regime, the system is in the following evaluated under two distinct nonlinear, second-order car-following models.

\subsection{The Intelligent Driver Model}
For our investigations, we employ two different nonlinear car-following models, starting with the Intelligent Driver Model (IDM) proposed by Treiber et al.~\cite{treiber2000congested}.
The IDM is a second-order car-following model given by 
% the position $x_n(t)$ and velocity $v_n(t)$ of the $n$-th vehicle at time $t$ evolve according to the dynamics:
% \begin{equation}
% \frac{d x_n(t)}{dt} = v_n(t),
% \end{equation}
% while
\begin{equation}
\frac{d v_n(t)}{dt} = a \left[1 - \left(\frac{g_n\big(v_n(t), v_{n+1}(t)\big)}{d_n(t) - \ell}\right)^2 - \left(\frac{v_n(t)}{v_0}\right)^4 \right],
\end{equation}
where $d_n(t)$ is the distance to the leading vehicle, $\ell=5$~m is the vehicle length, $v_0$ is the maximum (desired) speed, $a>0$ is the maximum acceleration, and $g(v_n, v_{n+1})$ is the desired gap given by
\begin{equation}
    g_n(v_n, v_{n+1}) = g_0 + v_n T_n - \frac{v_n (v_{n+1} - v_n)}{2 \sqrt{a b}},
\end{equation}
where $g_0=1$~m is the minimum gap, $T_n>0$ is the desired time gap that can be specific to each vehicle, and $b>0$ is the comfortable deceleration.

\subsection{The Full Velocity Difference Model}
The second model we consider is the Full Velocity Difference Model (FVDM) proposed by Jiang et al.~\cite{jiang2001full}. The FVDM is a relaxation-driven second-order model given by
% in which the position $x_n(t)$ and velocity $v_n(t)$ of the $n$-th vehicle at time $t$ evolve according to the dynamics:
% \begin{equation}
%     \frac{d x_n(t)}{dt} = v_n(t),
% \end{equation}
% while the acceleration is governed by:
\begin{equation}
    \frac{d v_n(t)}{dt} = \frac{1}{\tau_1} \Big[ V_n\big(d_n(t)-g_0\big) - v_n(t) \Big] + \frac{1}{\tau_2} \big( v_{n+1}(t) - v_n(t) \big),
\end{equation}
where %$d_n(t)=x_{n+1}(t)-x_n(t)$ is the distance headway to the leading vehicle~($n+1$). The 
the parameter $\tau_1>0$~[s] is the relaxation time, dictating the time scale over which a driver adapts to their optimal target velocity, and $\tau_2>0$~[s] is the relative velocity response time. The optimal velocity function $V_n(\cdot)$ prescribes the desired equilibrium speed for a given spacing and is conventionally defined as the piecewise linear function \cite{newell2002simplified}
\begin{equation}
    V_n(d)=\min\big\{v_0,\max\{0,(d-\ell)/T_n\}\big\}
\end{equation}
where, as previously, $v_0$ is the maximum speed, $\ell$ is the vehicle length including a stopping distance, and $T_n>0$ is the desired time gap specific to each vehicle.

Since the FVDM has no collision-avoidance mechanism, vehicles can overlap in fully developed waves, which would artificially increase the mean speed. In the time-domain simulations, each updated speed is therefore limited to the safe speed $v_n\le -b_{\max}\Delta t+\sqrt{(b_{\max}\Delta t)^2+v_{n+1}^2+2b_{\max}(d_n-\ell)}$~\cite{krauss1998microscopic}, with $b_{\max}=8$\,m/s$^2$ and time step $\Delta t=0.02$\,s. This constraint is inactive at and near equilibrium and therefore does not affect the linear stability analysis.

\subsection{Implementation of the Cooperative Buffering Strategy}

A periodic ring-road topology of length $L = 201$ meters populated by $N = 20$ vehicles similar to the setup of Sugiyama et al.\ (2008)~\cite{sugiyama2008traffic} is considered. This spatial configuration establishes a capacity-constrained, dense traffic regime inherently susceptible to the spontaneous emergence of SGWs. 
To operationalize the theoretically optimal cooperative buffering strategy within these nonlinear frameworks, a dimensionless buffering parameter $B \ge 1$ is introduced. For the single designated buffer vehicle, the nominal time gap $T>0$ is amplified such that $T_{\text{buffer}} = B \cdot T$. Crucially, to rigorously isolate the stabilizing effect of spatial non-uniformity from trivial density reductions, we enforce a strict mean-preserving transformation. The remaining $N-1$ platooning vehicles cooperatively compress their respective time gaps to
\begin{equation}
T_{\text{platoon}} = \frac{N - B}{N - 1} T.
\label{platoon}
\end{equation}
Consequently, the system-wide average time gap is strictly maintained at nominal time gap $T$. The traditional uniform traffic paradigm corresponds to the homogeneous baseline state of $B = 1$. 
The analytical results confirm that the single-buffer solution is optimal under the linear, long-wavelength assumptions. However, multi-buffer configurations are more feasible in practice: the required buffering time is distributed across several buffers (and thus reduced per vehicle).
Multiple buffers also mitigate intra-platoon instabilities that can arise when platoons become too large. 
More generally, for $K>1$ buffer vehicles each holding $T_{\text{buffer}} = B\,T$, the mean-preserving platoon time gap reads
\begin{equation}
T_{\text{platoon}} = \frac{N - KB}{N - K} T.
\label{platoonK}
\end{equation}
Equations~\eqref{platoon} and~\eqref{platoonK} are subject to explicit feasibility constraints. Positivity of the platoon time gap requires $KB<N$, and requiring the platooning vehicles to respect a minimum admissible time gap $T_{\text{min}}$ further imposes $T_{\text{platoon}}> T_{\text{min}}$. We use $T_{\text{min}}=0.6$~s, the shortest time gap tested in the CACC field experiments of~\cite{milanes2013cooperative}. The constraint yields $B<B_{\max}=\big(N-(N-K)T_{\text{min}}/T\big)/K$, i.e., $B_{\max}=8.6$, $4.6$, and $3.27$ for $K=1,2,3$ and $T=1$~s. The simulated configurations have $T_{\text{platoon}}=0.74$, $0.68$, $0.78$, and $0.65$~s for $(K,B)=(1,6)$, $(1,7)$, $(2,3)$, and $(3,3)$, and the sweep in Fig.~2 ($K=1$ and $B\le8$) satisfies the constraint. The single buffer with $B=9$ in Fig.~4(b) ($T_{\text{platoon}}=0.58$~s) lies outside it and only illustrates intra-platoon instability. The uniform baseline corresponds to a fleet of identical constant-time-gap controllers, the spacing policy of ACC systems (ISO 15622), whereas the JAD baseline introduced below is a simplified, static representation of the gap-holding idea behind JAD, in which one vehicle enlarges its gap without compression by the followers.

\subsection{Numerical Experiments}

In the following, we first numerically analyze in Section~\ref{sec:eigenvalues} the eigenvalue spectrum of the linearized system to assess the stabilizing effect of the buffer strategy on string stability across all wavelengths. Second, we evaluate the resulting traffic flow gains in Section~\ref{sec:MacroOpt}. Finally, we address practical deployment constraints in Section~\ref{Sec:multibuff} by extending the framework to a multi-buffer strategy tailored for large-scale vehicle platoons. Complementary comfort and smoothness indicators, evaluated for two distinct initial conditions, are reported in Section \ref{sec:confort}. Throughout this section, we use the following terminology: on a closed ring of fixed density $\rho = N/L$, the stationary flow is $Q = \rho\, \bar v$, so that maximising the equilibrium mean speed $\bar v$ and maximising the flow $Q$ (used synonymously with throughput hereafter) coincide. Capacity denotes the maximum flow achievable under the string-stability requirement. We note that these quantities, while directly linked in the closed-ring setting, are not generally interchangeable in open traffic systems.

\subsubsection{Eigenvalue Analysis and Expansion of the String-Stable Regime\label{sec:eigenvalues}}

  \begin{figure}[htbp]
      \centering

      \begin{subfigure}[t]{0.48\textwidth}\centering \includegraphics[width=\linewidth]{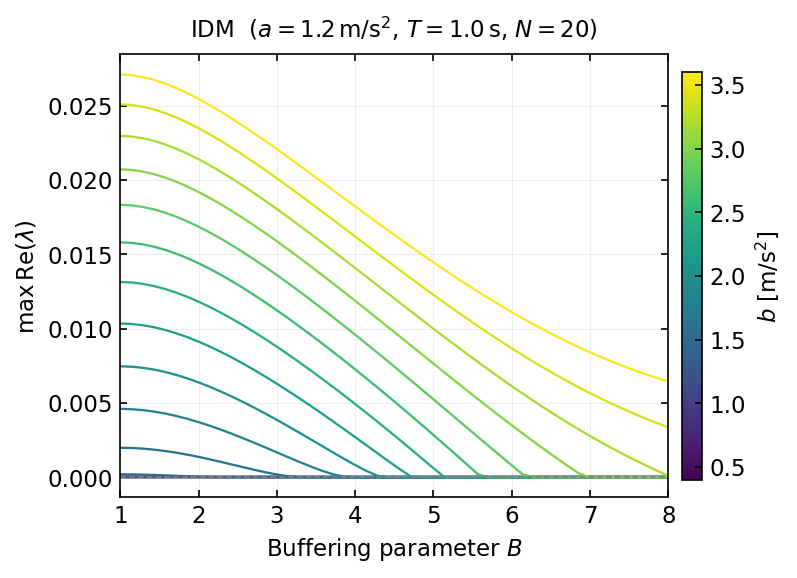}\caption{}\label{fig:re_a}\end{subfigure}
      \hfill
      \begin{subfigure}[t]{0.48\textwidth}\centering \includegraphics[width=\linewidth]{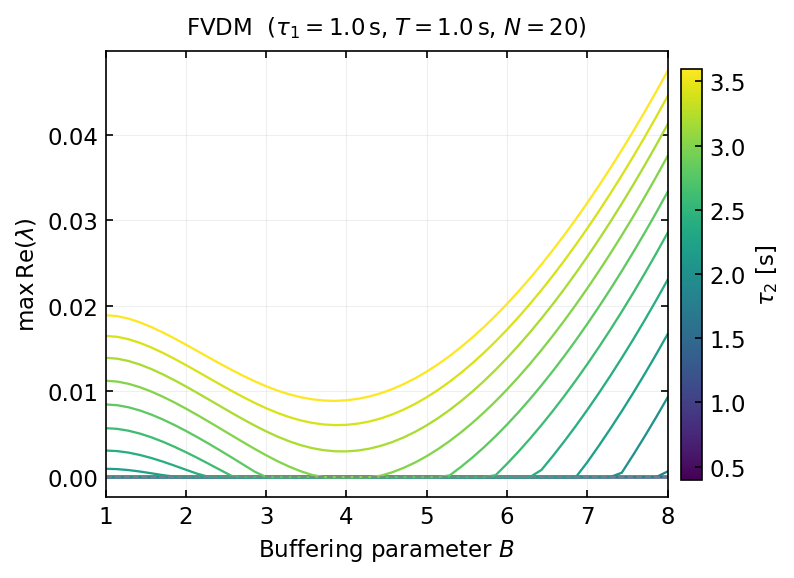}\caption{}\label{fig:re_b}\end{subfigure}

      \caption{Maximum real part of the dominant eigenvalue, $\max Re(\lambda)$, as a function of the buffering parameter $B$ for a platoon of $N = 20$ vehicles. Each curve corresponds to a different value of the respective sensitivity parameter (color-coded). The dashed line at max $Re(\lambda) = 0$ marks the string-stability boundary. (a) IDM: increasing $b$ raises the instability growth rate, while increasing $B$ monotonically reduces it, restoring string stability for $b\le3.2$\,m/s$^2$. (b) FVDM: a non-monotonic dependence on $B$ is observed; moderate buffering stabilises the platoon, but excessive buffering reintroduces instability, particularly for large values of $\tau_2$.}
      \label{fig:eigen_analysis}
  \end{figure}

Fig.~\ref{fig:re_a} presents the stability spectrum for the IDM with $N = 20$ vehicles, desired velocity $v_0 = 20\;\text{m/s}$, nominal time gap $T = 1.0\;\text{s}$, and maximum acceleration $a = 1.2\;\text{m/s}^2$, as the comfortable deceleration $b$ is swept from $0.5$ to $3.4\;\text{m/s}^2$.
Under uniform conditions ($B = 1$), the system is categorically unstable across the entire parameter range tested for $b$, i.e., $\max\operatorname{Re}(\lambda) > 0$, where $\lambda$ are the linearised system eigenvalues. % for every tested value of $b$.  
Critically, the instability growth rate scales with $b$. Configurations with aggressive deceleration preferences exhibit the largest positive system eigenvalues, consistent with the IDM's well-documented sensitivity of string stability to the $a/b$ ratio~\cite{treiber2012traffic}.

As the buffering parameter~$B$ increases beyond unity, the dominant eigenvalue's real part decreases for every value of~$b$. This monotonicity is a direct consequence of the IDM's acceleration-driven instability mechanism. Enlarging the buffer vehicle's time gap by a factor of $B$ increases the dissipation margin available to absorb upstream perturbations, while the $(N-1)$ platooning vehicles retain sufficient headway to remain individually well-posed. For each $b$, there exists a finite critical $B^*$ beyond which $\max\operatorname{Re}(\lambda) \leq 0$. This threshold is lowest for small $b$ and highest for large $b$, confirming that the cooperative strategy's stabilization burden is proportional to the system's intrinsic instability.

Fig.~\ref{fig:re_b} displays the corresponding analysis for the FVDM with $\tau_1 = 1.0\;\text{s}$ and $T = 1.0\;\text{s}$, sweeping the relative velocity sensitivity $\tau_2$ over the range $[0.5,3.5]$~s.  The qualitative picture differs from the IDM.  Moderate cooperative buffering ($1 < B < 6$) effectively suppresses the dominant long-wavelength instabilities, driving $\max\operatorname{Re}(\lambda)$ toward zero.  However, for $B > 6$ and large $\tau_2$, the eigenvalue spectrum exhibits a \emph{resurgence}, where $\max\operatorname{Re}(\lambda)$ increases again, and the system re-enters the unstable regime. The physical mechanism underlying this non-monotonicity is the FVDM's relaxation-driven instability. As $B$ grows, the $(N-1)$ follower vehicles must operate at progressively compressed headways to conserve the total road space. When these headways approach the model's intrinsic relaxation limit, high-frequency, short-wavelength oscillation modes become excited---modes that are invisible to the longest-wavelength stability condition~\eqref{eq:general_stability} but are captured by the full spectral analysis.  The result is an \emph{optimal} buffering level, $B^{\mathrm{opt}} \approx 3$--$4$, that balances long-wavelength damping against short-wavelength excitation.  Even at this optimum, the cooperative configuration vastly outperforms the uniform baseline, confirming that the stabilizing mechanism is robust across relaxation-driven models.

\subsubsection{Macroscopic Traffic Flow Optimization\label{sec:MacroOpt}}
%%%%% IN this chapter Distinction to JAD must be introduced!
Having established that cooperative buffering expands the string stable parameter space, this section quantifies the direct impact on traffic efficiency. For each model, parameter value, and allocation $(K,B)$, all time gaps are multiplied by a common factor $\eta\ge1$, which lowers mean speed $\bar v$ and thus flow $Q$ since the density is constant, but has a stabilising effect. The maximum string-stable flow is therefore the equilibrium flow at the smallest factor $\eta^*$ for which the linearised ring ($N=20$, $L=201$\,m) is string-stable for all wavelengths, determined by bisection.
If $\eta^*=1$, the configuration with $T=1$\,s is already stable. 
%Therefore, the optimal flow is bounded by $\big(1-(\ell+g_0) N/L\big)/T\approx 1451$~veh/h for the IDM and $(1-\ell N/L)/T\approx 1809$~veh/h for the FVDM.

\begin{figure}[htbp]
    \centering

    \begin{subfigure}[t]{0.48\textwidth}\centering \includegraphics[width=\linewidth]{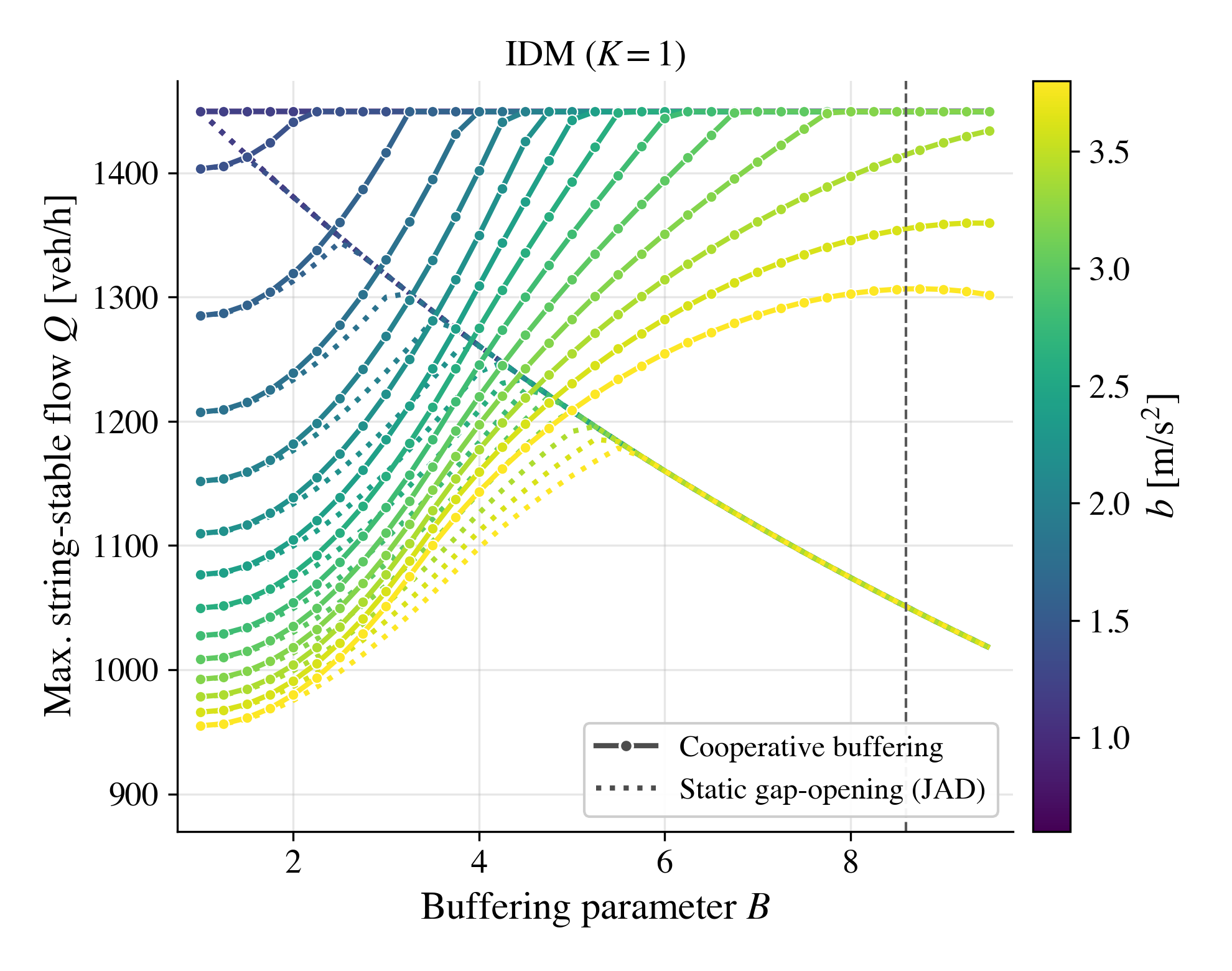}\caption{}\label{fig:flow_a}\end{subfigure}
    \hfill 
    \begin{subfigure}[t]{0.48\textwidth}\centering \includegraphics[width=\linewidth]{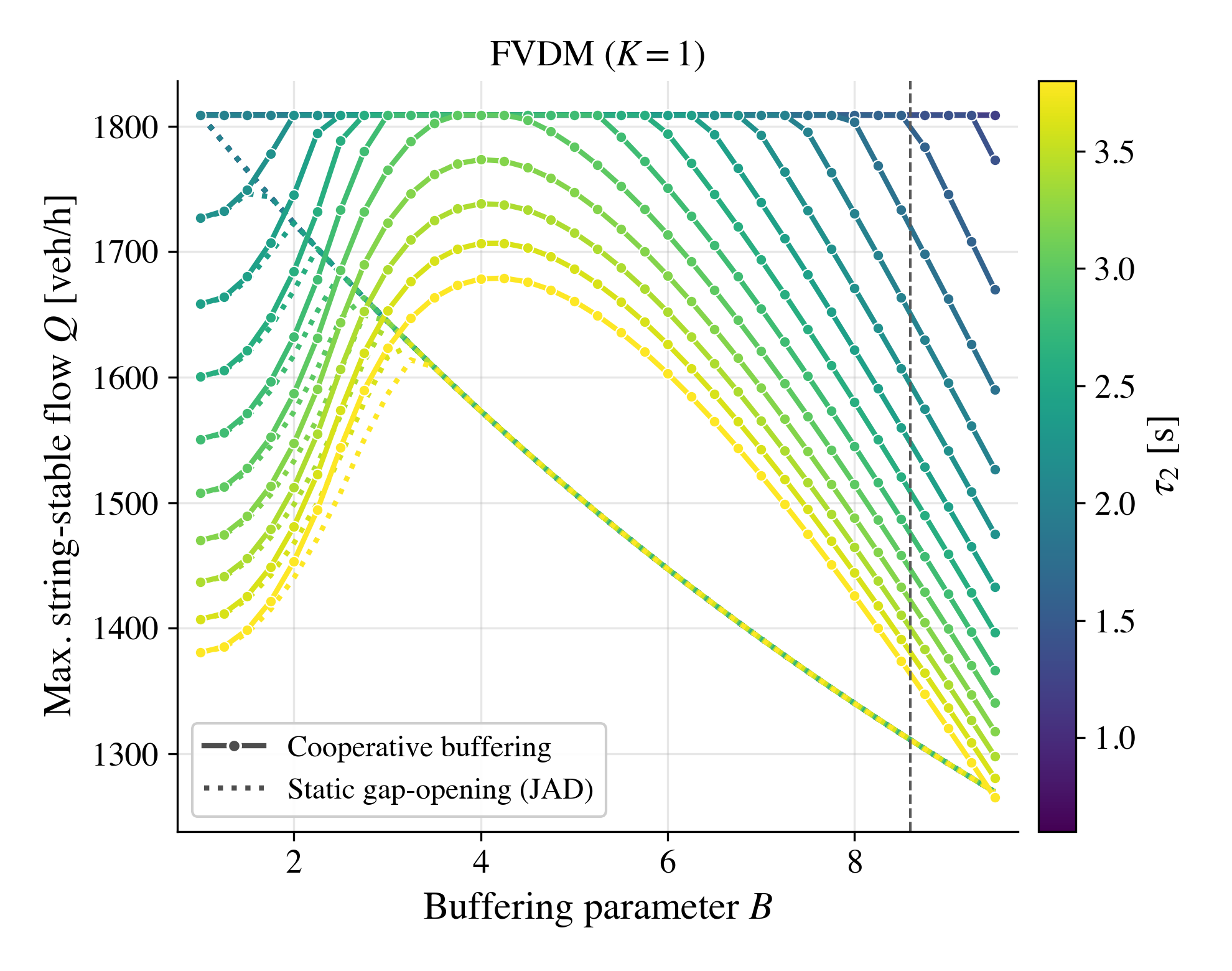}\caption{}\label{fig:flow_b}\end{subfigure}

    \caption{Maximum string-stable traffic flow $Q$ [veh/h] as a function of the buffering parameter $B$ for a single buffer vehicle ($K=1$). For each allocation, $Q$ is the largest equilibrium flow of the ring ($N=20$, $L=201$\,m) at which the allocation is linearly string-stable for all wavelengths ($\max \text{Re}(\lambda) \le 0$), obtained by scaling all time gaps by the smallest stable common factor $\eta^*\ge1$ (Section~IV-D2). Solid lines: cooperative buffering; dotted lines: static gap-opening baseline (JAD). Dashed line: feasibility limit $B_{\max}=8.6$ ($T_{\text{platoon}}<T_{\text{min}}$ for $B>B_{\max}$). (a)~IDM: Evaluated across varying comfortable deceleration parameters $b$. For highly unstable configurations (high $b$), the uniform baseline ($B=1$) suffers severe capacity drops. Increasing $B$ effectively dissipates perturbations and restores the equilibrium flow of the simulated configuration ($1449$\,veh/h) for $b\le3.2$\,m/s$^2$. (b)~FVDM: Evaluated across varying relative velocity sensitivities $\tau_2$. The maximum string-stable flow exhibits a non-monotonic dependence on $B$. For $\tau_2\le3.0$\,s, the equilibrium flow ($1809$\,veh/h) is attained within a window of moderate buffers that narrows with increasing $\tau_2$ (e.g., $3\le B\le5.25$ for $\tau_2=2.8$\,s); for larger $B$, intra-platoon instability reduces the flow again. For $\tau_2\ge3.2$\,s, it is not attained. In both models, an adequately dimensioned cooperative buffering outperforms the uniform and the gap-opening baseline.}
    \label{fig:flow} 
\end{figure}

Figure~\ref{fig:flow_a} presents the traffic flow $Q$ in veh/h for the IDM as a function of the buffering parameter~$B$, with curves color-coded by the comfortable deceleration parameter~$b$. At the uniform baseline ($B=1$), the achievable throughput is severely constrained. This applies especially for large values of~$b$, where SGW are most prevalent and cause a decrease of the average flow. As~$B$ increases, the flow rises monotonically and substantially. For high comfort parameters ($b\geq 2.5$\,m/s$^2$), the cooperative strategy unlocks high throughput gains, demonstrating
that the single buffer vehicle absorbs the system's entire stability burden and frees the remaining $N-1$ vehicles to operate at higher-density, higher-flow equilibria that would be categorically unstable under uniform
conditions. The JAD baseline (dotted lines) exhibits a strikingly different trajectory.  At $B = 1$, JAD and cooperative buffering coincide by construction. For $1<B\le B_{\max}$, however, the JAD throughput never exceeds that of cooperative buffering and, once the flow is stabilised, decreases with $B$. Although JAD can successfully stabilize traffic, increasing the time gap of a single vehicle without cooperatively compressing the follower vehicles forces a reduction in the mean equilibrium speed. Consequently, overall macroscopic traffic flow is reduced. The larger the buffer parameter $B$, the more drastically the flow drops, following an approximately linear trend.

Figure~\ref{fig:flow_b} displays the analogous analysis for FVDM, with curves parametrized by the relative velocity sensitivity~$\tau_2$. Consistent with
the non-monotonic spectral behaviour identified in Figure~\ref{fig:re_b}, the maximum stable flow initially increases with~$B$ but saturates and, for large~$\tau_2$, decreases again at excessive buffering levels.
This saturation reflects the FVDM's relaxation dynamics. As the buffer gap grows, the compressed platoon headways approach the model's minimum
relaxation threshold, exciting short-wavelength instabilities that cap further capacity gains. The flow gain over the uniform baseline grows with $\tau_2$ and reaches about $20$\,\% for $\tau_2\ge3.0$\,s, although the equilibrium flow is no longer attained for $\tau_2\ge3.2$\,s. For moderate buffering ($B\approx 3$--$5$), the cooperative strategy yields a significant increase in throughput.
Once stabilised, the JAD baseline again decreases with $B$, following an approximately linear decline that converges across all $\tau_2$ values.  This convergence arises because JAD's throughput penalty is governed purely by the inflated mean headway. %, which is independent of the model's stability parameters.

\subsubsection{Mitigating Intra-Platoon Instability via Multi-Buffer Configurations\label{Sec:multibuff}}

\begin{figure}[htbp]
      \centering

      \begin{subfigure}[t]{0.48\textwidth}\centering \includegraphics[width=\linewidth]{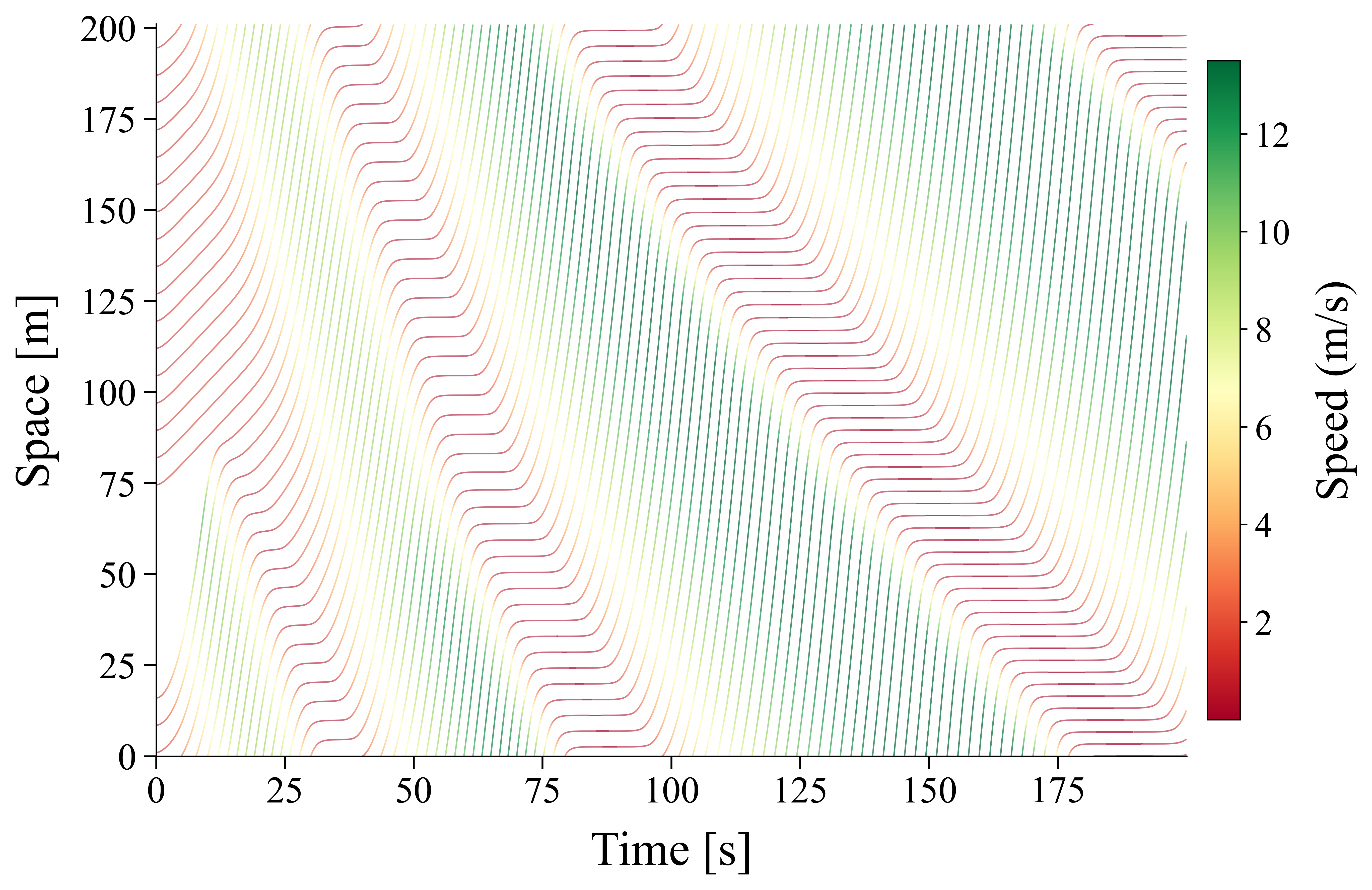}\caption{}\label{fig:ts_a}\end{subfigure}
      \hfill
      \begin{subfigure}[t]{0.48\textwidth}\centering \includegraphics[width=\linewidth]{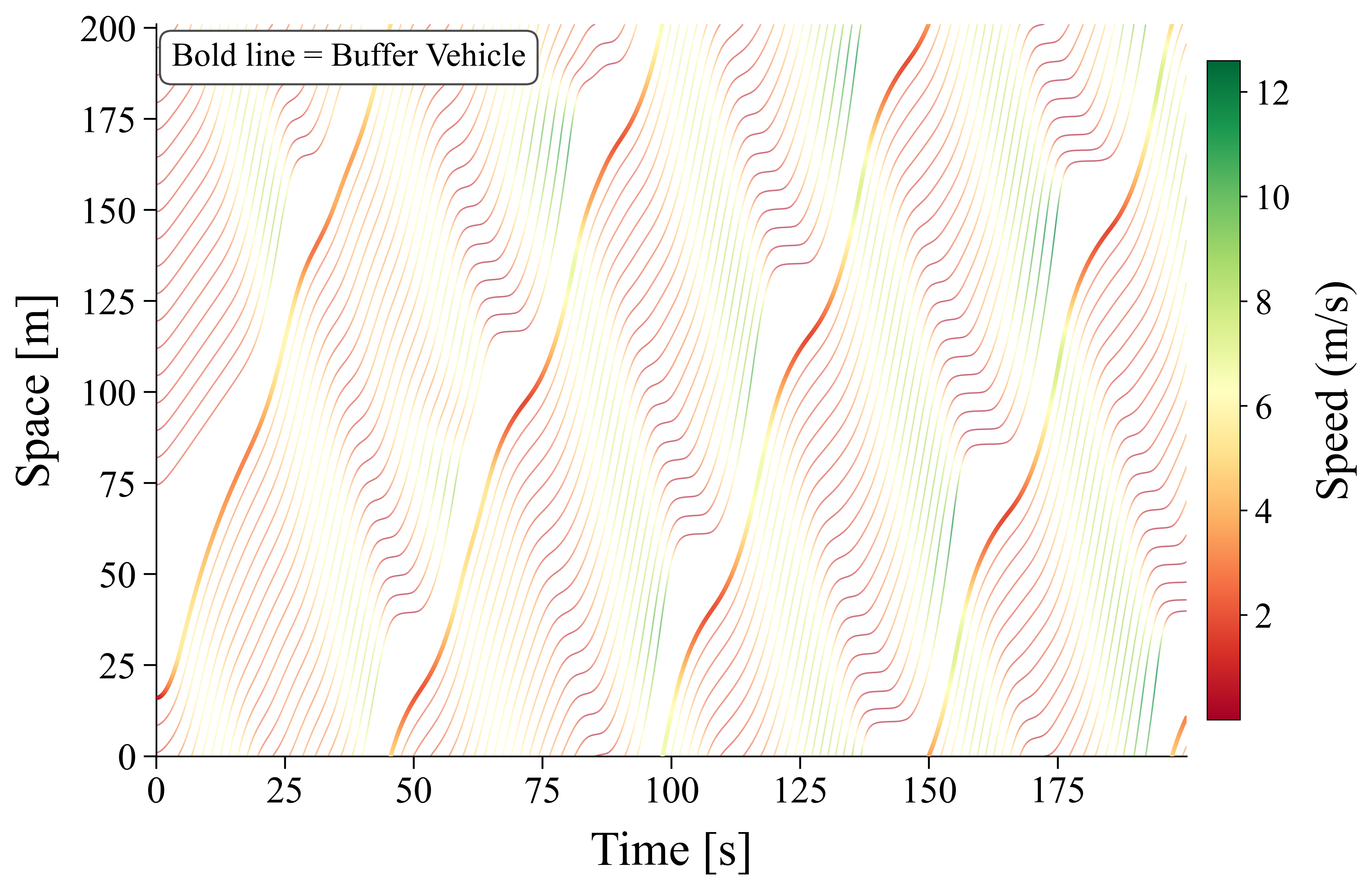}\caption{}\label{fig:ts_b}\end{subfigure}

      \vspace{0.5em}

      \begin{subfigure}[t]{0.48\textwidth}\centering \includegraphics[width=\linewidth]{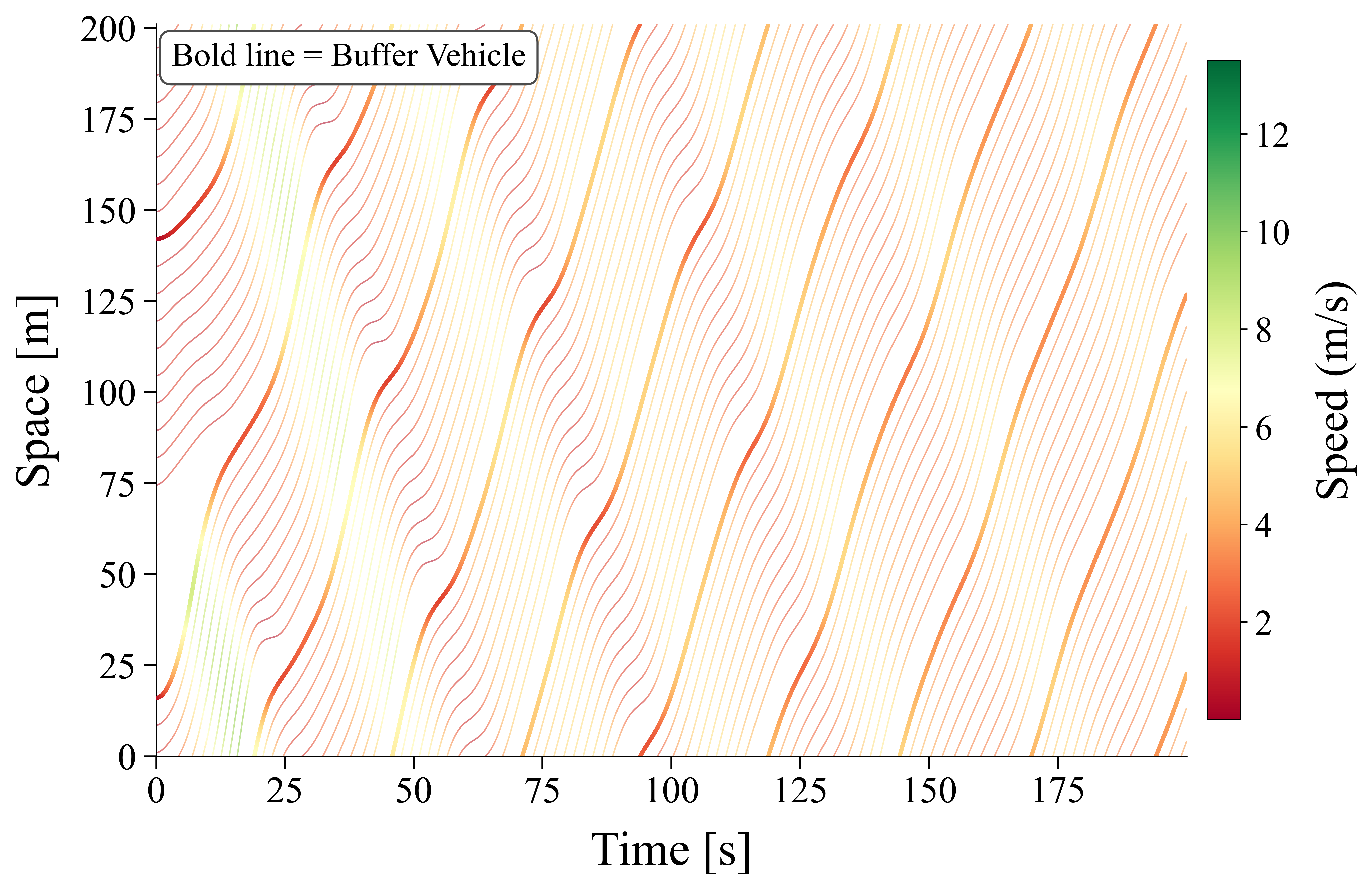}\caption{}\label{fig:ts_c}\end{subfigure}
      \hfill
      \begin{subfigure}[t]{0.48\textwidth}\centering \includegraphics[width=\linewidth]{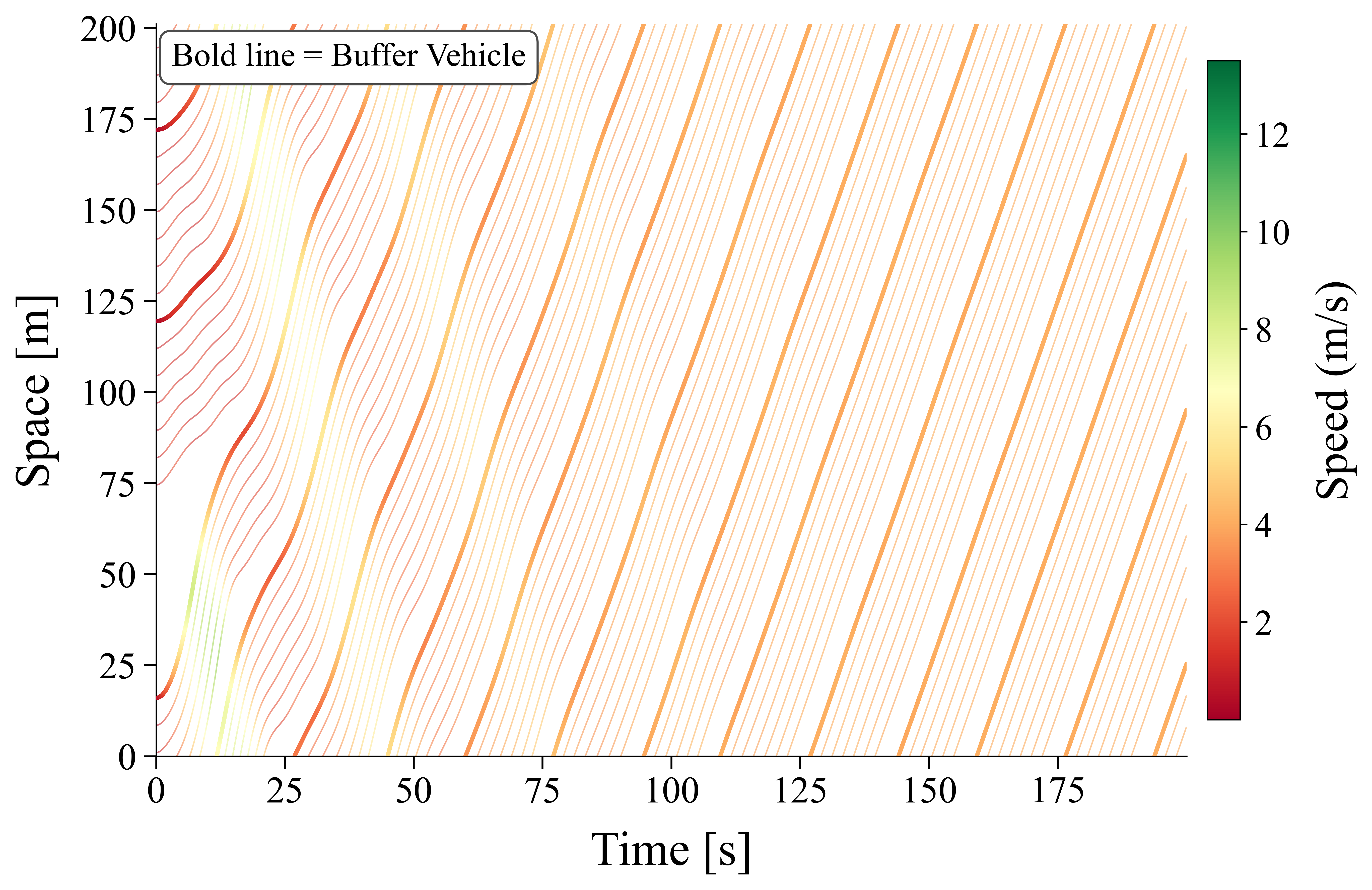}\caption{}\label{fig:ts_d}\end{subfigure}

      \caption{Time--space diagrams of the FVDM ring-road simulations under different buffer-vehicle configurations with jam initial condition. Vehicle trajectories are coloured by their instantaneous speed (red: slow, green: fast) and
  the bold line marks the buffer vehicle(s). (a)~Unstable uniform flow. SGW propagate upstream and persist over the entire observation window. (b)~Single-buffer configuration with $B = 9$ and intra-platoon instability. (c)~Multi-buffer configuration with
   two buffer vehicles each keeping $B = 3$. SGW dissipate partly.  (d)~Multi-buffer configuration with three buffer vehicles each keeping $B = 3$. SGW dissipate completely and the mean speed is improved. The trajectories can be computed and visualized online; see~\cite{SimMultiBufferingStrategy}.}
      \label{fig:time_space_comparison}
  \end{figure}
  
%While the results in the previous sections indicate that a single buffer vehicle is an optimal theoretical solution for mitigating SGWs by preserving traffic flow stability,

Mitigating SGWs with a single vehicle exhibits practical limitations when the platoon size~$N$ becomes very large. As the mathematically optimal solution scales towards a long platoon, the required time gap for the single buffer vehicle becomes excessively large. This presents two major real-world challenges that motivate the transition to a multi-buffer strategy. First, unavoidable cut-ins: In highway environments, maintaining a massive gap in front of a single vehicle inevitably invites human-driven vehicles from adjacent lanes to cut in~\cite{hayat2025traffic}. Such uncontrolled merging manoeuvres would instantly compromise the buffer gap and introduce severe new perturbations into the system. Second, intra-platoon instability: Even if cut-ins are prevented, an extremely long, tightly packed continuous string of follower vehicles is highly susceptible to internal string instability. Small reaction-time delays can accumulate within a long platoon, exciting short-wavelength oscillations that amplify. 

This is precisely the phenomenon observed for the FVDM in Fig.~\ref{fig:re_b} when the buffer parameter is set too high. These effects are directly visible in the time--space diagrams of Fig.~\ref{fig:time_space_comparison}. Fig.~\ref{fig:ts_a} shows the uniform FVDM solution ($B=1$), where SGWs are clearly present and propagate upstream throughout the entire observation window. Fig.~\ref{fig:ts_b} shows the single-buffer configuration with $B=9$. The buffer vehicle stabilizes the traffic and exhibits less speed variation compared to the uniform case, but instability persists within the compressed follower platoon. A natural remedy is to reduce the number of vehicles per sub-platoon by distributing the buffering effort across multiple buffer vehicles. This is demonstrated in Figs.~\ref{fig:ts_c} and \ref{fig:ts_d}, where two and three vehicles each maintain a buffer of $B=3$. While the single-buffer solution is unable to stabilize the flow entirely, the multi-buffer configuration succeeds. A parametrisable simulation platform is available online; see~\cite{SimMultiBufferingStrategy}.

 \begin{figure}[htbp]
      \centering

      \begin{subfigure}[t]{0.48\textwidth}\centering \includegraphics[width=\linewidth]{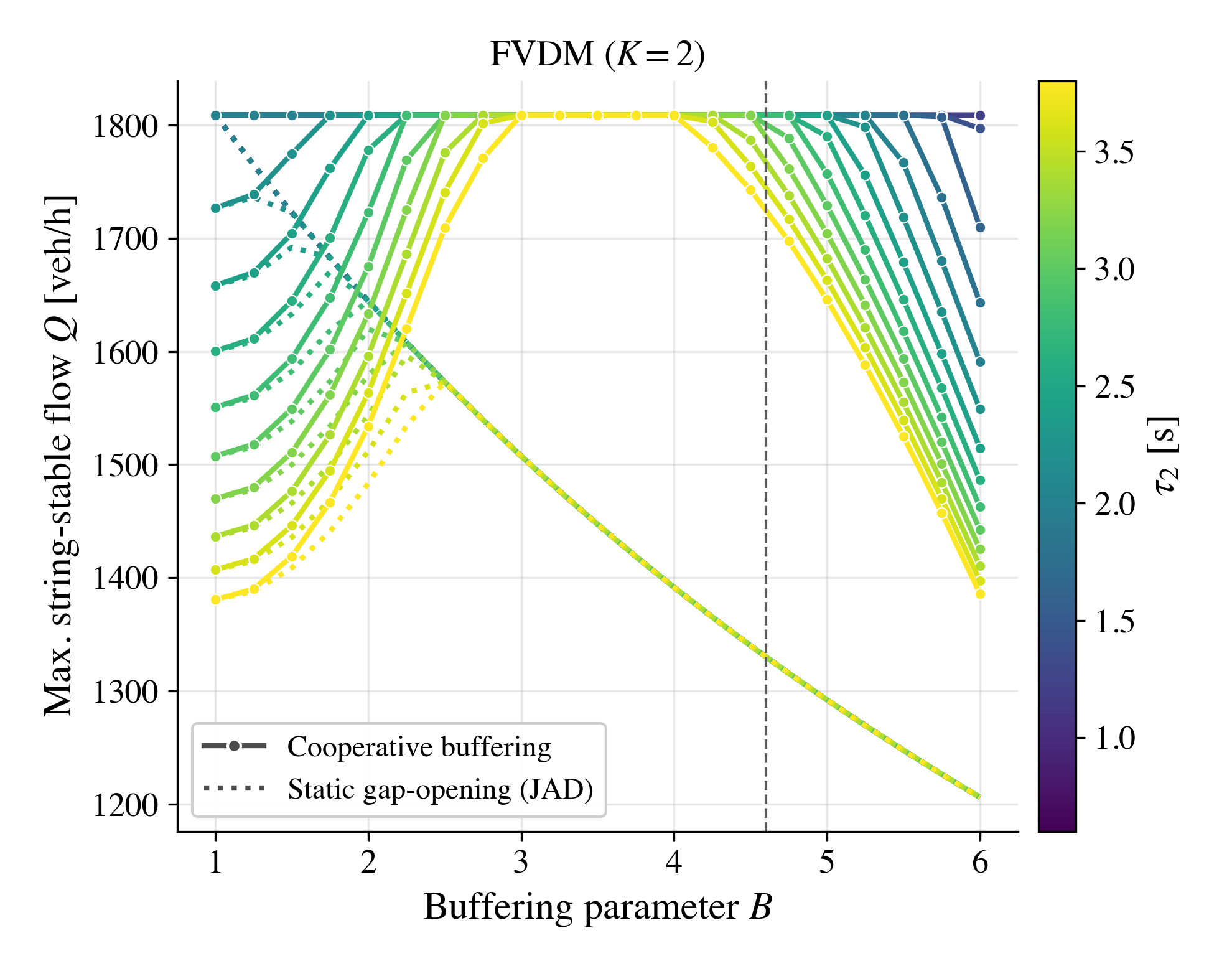}\caption{}\label{fig:multibuf_a}\end{subfigure}
      \hfill
      \begin{subfigure}[t]{0.48\textwidth}\centering \includegraphics[width=\linewidth]{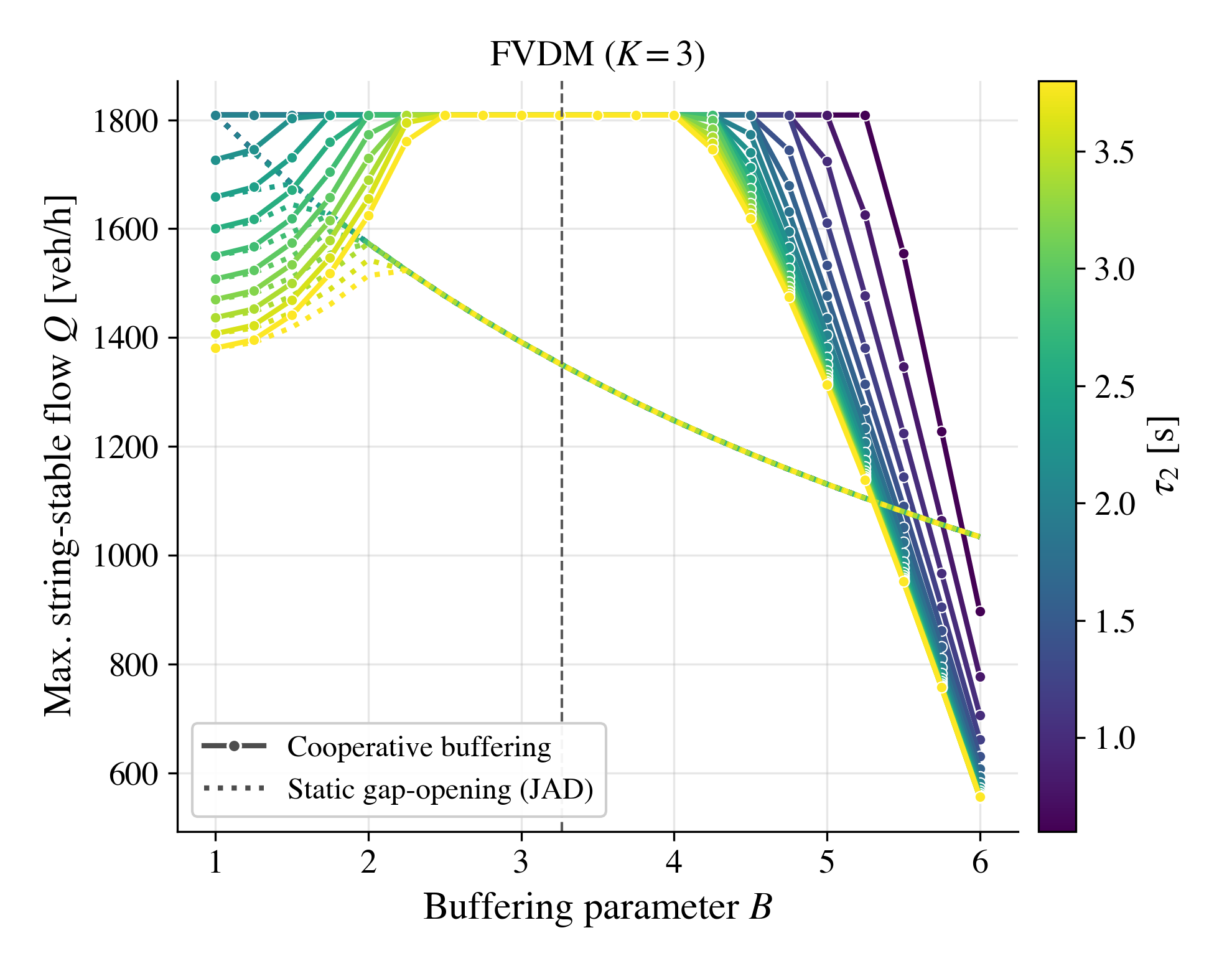}\caption{}\label{fig:multibuf_b}\end{subfigure}

     \caption{Maximum string-stable traffic flow $Q$ [veh/h] as a function of the buffering parameter $B$ for the FVDM with multiple buffer vehicles, computed as in Fig.~\ref{fig:flow}. Curves are evaluated across varying relative velocity sensitivities $\tau_2$. Solid lines: cooperative buffering; dotted lines: static gap-opening baseline with $K$ gap-opening vehicles. Dashed lines: feasibility limit $B_{\max}(K)$. (a)~$K=2$ buffer vehicles ($B_{\max}=4.6$): The cooperative buffering allocation attains the equilibrium flow ($1809$\,veh/h) for all tested $\tau_2$ (e.g., $2.25\le B\le4.5$ for $\tau_2=2.8$\,s). (b)~$K=3$ buffer vehicles ($B_{\max}=3.27$): The equilibrium flow is attained for all tested $\tau_2$ from $B=2.5$ up to $B_{\max}$, which includes the configuration $K=3$, $B=3$ of Table~I. Well beyond $B_{\max}$ ($B>4$), the flow drops sharply due to the very short platoon time gaps. For $B>1$, the gap-opening baseline remains below the equilibrium flow in both configurations, and the cooperative buffer dominates it across all tested $\tau_2$ within the feasible range.}
      \label{fig:multibuffer_comparison}
  \end{figure}
  
To systematically quantify the stabilizing effect of the multi-buffer approach, the macroscopic flow analysis from Fig.~\ref{fig:flow_b} is extended to two buffer vehicles in Fig.~\ref{fig:multibuf_a} and three buffer vehicles in Fig.~\ref{fig:multibuf_b}. As observed previously, the single-buffer FVDM fails to guarantee string stability for large values of the sensitivity parameter $\tau_2$. By transitioning to a multi-buffer architecture, we overcome the challenge of intra-platoon instability. This strategy fully stabilizes the traffic flow and sustains the equilibrium flow for all tested $\tau_2$, e.g., for $2.5\le B\le B_{\max}=3.27$ with $K=3$, which includes the configuration $K=3$, $B=3$ of Section~\ref{sec:confort}. The admissible range of $B$, however, shrinks with $K$. %, confirming its viability for large-scale, real-world deployment.

\subsubsection{Comfort, Smoothness, and Safety Indicators\label{sec:confort}}
The stability and efficiency metrics used so far, namely $\max \text{Re}(\lambda)$ and the maximum string-stable flow $Q$, do not directly measure driving comfort or smoothness. This section, therefore, reports the comfort indicators: the standard deviation of speed ${\sigma}_v$, the standard deviation of acceleration ${\sigma}_a$, the root-mean-square jerk, and, as a safety indicator, the minimum net (bumper-to-bumper) gap $s_{min}$. These metrics are evaluated over the transient window $0$--$100$\,s and in the stationary long-time state (final $300$\,s of a $3000$\,s horizon, long enough for the slowly growing instabilities near the stability boundary to saturate).
Two initial conditions are considered: (i) the compact stopped jam used in the preceding experiments, and (ii) a uniform equilibrium flow perturbed by small Gaussian speed fluctuations ($\sigma=0.3$\,m/s), thereby complementing the jam-based scenarios of Figs. \ref{fig_strategy} and \ref{fig:time_space_comparison}.

Table \ref{tab:comfort} shows that, in the stationary state, in every scenario in which the uniform configuration develops sustained SGWs, adequately dimensioned cooperative buffering suppresses the speed and acceleration standard deviations and the jerk, while simultaneously increasing (IDM) or preserving (FVDM) the mean speed and enlarging the minimum gap. During the transient after a jam start, however, the minimum gap of the compressed platoon is smaller than in the uniform case. It occurs at the platoon vehicle directly ahead of the buffer, because the braking wave that forms behind the slowly accelerating buffer is amplified along the compressed platoon before the buffer absorbs it. In linearly unstable settings (IDM $b=2.5$ and $3.0$, FVDM $\tau_2=2.5$ and 2.8), sustained waves emerge from both initial conditions and converge to the same stationary state. The two initialisations differ only in how quickly this stationary state is reached.

\begin{table}[htbp]
\caption{Comfort and safety indicators, computed across all vehicles over the transient window and in the stationary state. Transient entries are reported as pairs \textbf{jam}\,/\,\textbf{pert}.\ for the jam and the perturbed-uniform initialisation. Both initialisations reach the same stationary state. The simulations of the FVDM use the collision-free speed constraint of Section~IV-B.}
\label{tab:comfort}
\centering
\resizebox{\textwidth}{!}{%
\begin{tabular}{llcccccccccc}
\hline
 & & \multicolumn{5}{c}{Transient ($0$--$100$\,s)} & \multicolumn{5}{c}{Stationary} \\
\cline{3-7}\cline{8-12}
Model & Configuration & $\bar v$ & $\sigma_v$ & $\sigma_a$ & Jerk & $s_\text{min}$ & $\bar v$ & $\sigma_v$ & $\sigma_a$ & Jerk & $s_\text{min}$ \\
 & & [m/s] & [m/s] & [m/s$^2$] & [m/s$^3$] & [m] & [m/s] & [m/s] & [m/s$^2$] & [m/s$^3$] & [m] \\
\hline
IDM ($b=2.5$) & Uniform ($B{=}1$) & 3.49\,/\,4.04 & 2.06\,/\,0.09 & 0.58\,/\,0.04 & 0.23\,/\,0.03 & 1.9\,/\,4.6 & 3.76 & 1.89 & 0.56 & 0.21 & 2.2 \\
IDM ($b=2.5$) & Single buffer ($K{=}1$, $B{=}6$) & 3.69\,/\,4.05 & 1.63\,/\,0.04 & 0.51\,/\,0.03 & 0.25\,/\,0.03 & 1.3\,/\,3.6 & 4.05 & $0.00$ & $0.00$ & $0.00$ & 4.0 \\
\hline
IDM ($b=3.0$) & Uniform ($B{=}1$) & 3.45\,/\,4.04 & 2.13\,/\,0.14 & 0.63\,/\,0.05 & 0.27\,/\,0.03 & 1.8\,/\,4.6 & 3.70 & 2.08 & 0.64 & 0.26 & 1.8 \\
IDM ($b=3.0$) & Single buffer ($K{=}1$, $B{=}6$) & 3.65\,/\,4.04 & 1.69\,/\,0.05 & 0.55\,/\,0.04 & 0.29\,/\,0.03 & 1.1\,/\,3.6 & 4.02 & 0.55 & 0.17 & 0.06 & 3.1 \\
IDM ($b=3.0$) & Single buffer ($K{=}1$, $B{=}7$) & 3.69\,/\,4.05 & 1.65\,/\,0.05 & 0.57\,/\,0.04 & 0.31\,/\,0.03 & 0.9\,/\,3.4 & 4.05 & $0.00$ & $0.00$ & $0.00$ & 3.8 \\
\hline
FVDM ($\tau_2=2.5$) & Uniform ($B{=}1$) & 4.89\,/\,5.05 & 3.60\,/\,0.04 & 1.49\,/\,0.03 & 6.38\,/\,0.03 & 0.0\,/\,4.7 & 4.99 & 3.47 & 1.27 & 7.77 & 0.0 \\
FVDM ($\tau_2=2.5$) & Single buffer ($K{=}1$, $B{=}6$) & 5.03\,/\,5.05 & 2.81\,/\,0.05 & 1.60\,/\,0.04 & 2.87\,/\,0.04 & 0.0\,/\,3.4 & 5.05 & $0.00$ & $0.00$ & $0.00$ & 3.7 \\
\hline
FVDM ($\tau_2=2.8$) & Uniform ($B{=}1$) & 4.86\,/\,5.05 & 3.74\,/\,0.05 & 1.66\,/\,0.04 & 5.98\,/\,0.03 & 0.0\,/\,4.7 & 4.96 & 3.82 & 1.51 & 7.39 & 0.0 \\
FVDM ($\tau_2=2.8$) & Single buffer ($K{=}1$, $B{=}6$) & 5.02\,/\,5.05 & 2.98\,/\,0.07 & 1.78\,/\,0.05 & 3.51\,/\,0.05 & 0.0\,/\,3.4 & 5.00 & 2.50 & 1.78 & 3.13 & 0.0 \\
FVDM ($\tau_2=2.8$) & Multi buffer ($K{=}3$, $B{=}3$) & 4.99\,/\,5.05 & 2.21\,/\,0.03 & 1.06\,/\,0.04 & 1.42\,/\,0.04 & 0.0\,/\,3.0 & 5.05 & $0.00$ & $0.00$ & $0.00$ & 3.3 \\
\hline
\end{tabular}}
\end{table}

The comfort-related model parameters, the maximum acceleration $a$ and comfortable deceleration $b$ in the IDM, and the relaxation times $\tau_1$ and $\tau_2$ in the FVDM, are precisely the parameters swept over broad ranges in the stability and flow analyses of Figs.~\ref{fig:eigen_analysis}--\ref{fig:multibuffer_comparison}, so the reported stabilisation holds for every comfort calibration, from smooth to aggressive behaviour. Table~\ref{tab:comfort} provides the direct comfort indicators at representative calibrations. Since suppressing SGWs replaces oscillatory driving by near-constant-speed driving, these indicators corroborate that the stability gains reported above are accompanied by, rather than traded against, smoother and more comfortable dynamics.

\section{Conclusion and Future Directions\label{sec:5}}

In this paper, we explore strategies to mitigate stop-and-go waves in traffic flow, drawing from both deep reinforcement learning and classical car-following models. Through training DRL agents in a ring road environment inspired by the Sugiyama experiment, we observe in an exploratory experiment the emergence of a cooperative buffering strategy: one vehicle maintains a significantly larger time gap to act as a buffer, while the others follow closely. Simulations implementing the strategy in the Intelligent Driver and Full Velocity Difference models show that it suppresses SGWs and increases the mean speed compared to non-cooperative configurations. This modelling framework thus helps address the two paradoxes identified in the introduction.
First,  the strategy avoids the efficiency loss associated with uniformly large time gaps by concentrating the buffering effort in a single vehicle. 
Second, the investigations indicate that stability need not come at the expense of comfort.  The stability condition constrains only the time gaps, namely$\sum_n T_n^2\ge \alpha_N$, while the comfort-related parameters of the underlying models (the maximum acceleration $a$ and comfortable deceleration $b$ in the IDM, and the relaxation times $\tau_1$ and $\tau_2$ in the FVDM) remain free calibration variables. Since the numerical evaluation sweeps precisely these parameters over broad ranges, the reported improvements hold across the entire range of comfort calibrations tested, from smooth to aggressive behaviour. 
Moreover, the smoothness indicators reported show that the strategy reduces the standard deviations of speed and acceleration as well as the jerk experienced by all vehicles.
However, maintaining a large headway for the buffer vehicle may prove difficult in practice  due, for instance, to cut-in manoeuvres in a multi-lane environment, and platooning can exhibit short-wave instabilities in extreme conditions. This motivates the development of multi-buffer solutions, for which the presented simulation results are promising to mitigate these short-wavelength instabilities.
%Furthermore, by formulating an optimization problem under linear stability constraints, we proved that this buffer approach leads to higher mean speed and overall traffic efficiency.

Generally speaking, future traffic systems could benefit from cooperative behaviour, potentially enabled by connected and autonomous vehicles. 
This strategy may be particularly relevant to truck platoons, where tight spacing among platooning vehicles is desirable for aerodynamic drag reduction, while a buffer vehicle could help preserve string stability.
Although current implementations face challenges such as the need for vehicle-to-vehicle communication, the study highlights the potential benefits of asymmetric cooperative strategies for future traffic control design.
Looking ahead, several avenues for extension warrant investigation. First, empirical validation through real-world experiments. Second, while the linear optimization proves that a single buffer vehicle is globally optimal for mitigating long-wavelength modes, the simulations reveal that multi-buffer configurations are practically necessary to handle short-wavelength intra-platoon instabilities. Future work should therefore investigate dynamically adjusting the number and position of buffer vehicles in real time. Third, systematic benchmarking against explicit ACC/CACC controller implementations and established traffic-smoothing controllers in open-road settings would further delineate the practical advantage of asymmetric time-gap allocation.

%tactical/strategical 
% cooperation: require connection

%Case where the cost function also depends on the speed of the predecessor?

\section*{Acknowledgment}
  The authors acknowledge the research project \emph{SmartACC} funded by the Deutsche Forschungsgemeinschaft (DFG, German Research Foundation), grant number 546728715.

%\tableofcontents

\bibliographystyle{IEEEtran}
\bibliography{biblio}

@article{newell2002simplified,
  title={A simplified car-following theory: a lower order model},
  author={Newell, Gordon Frank},
  journal={Transportation Research Part B: Methodological},
  volume={36},
  number={3},
  pages={195--205},
  year={2002},
  publisher={Elsevier}
}

@article{ciuffo2021requiem,
  title={Requiem on the positive effects of commercial adaptive cruise control on motorway traffic and recommendations for future automated driving systems},
  author={Ciuffo, Biagio and Mattas, Konstantinos and Makridis, Michail and Albano, Giovanni and Anesiadou, Aikaterini and He, Yinglong and Josvai, Szil{\'a}rd and Komnos, Dimitris and Pataki, Marton and Vass, Sandor and others},
  journal={Transportation Research Part C: Emerging Technologies},
  volume={130},
  pages={103305},
  year={2021},
  publisher={Elsevier}
}

@article{jiang2001full,
  title={Full velocity difference model for a car-following theory},
  author={Jiang, Rui and Wu, Qingsong and Zhu, Zuojin},
  journal={Physical Review E},
  volume={64},
  number={1},
  pages={017101},
  year={2001},
  publisher={APS}
}

@article{ngoduy2015effect,
  title={Effect of the car-following combinations on the instability of heterogeneous traffic flow},
  author={Ngoduy, Dong},
  journal={Transportmetrica B: transport dynamics},
  volume={3},
  number={1},
  pages={44--58},
  year={2015},
  publisher={Taylor \& Francis}
}

@article{li2014stop,
  title={Stop-and-go traffic analysis: Theoretical properties, environmental impacts and oscillation mitigation},
  author={Li, Xiaopeng and Cui, Jianxun and An, Shi and Parsafard, Mohsen},
  journal={Transportation Research Part B: Methodological},
  volume={70},
  pages={319--339},
  year={2014},
  publisher={Elsevier}
}

@article{korbmacher2025understanding,
  title={Understanding Collective Stability of {ACC} Systems: From Theory to Real-World Observations},
  author={Korbmacher, Raphael and Khound, Parthib and Tordeux, Antoine},
  journal={arXiv preprint arXiv:2504.04530},
  year={2025}
}

@article{franccois2018introduction,
  title={An introduction to deep reinforcement learning},
  author={Fran{\c{c}}ois-Lavet, Vincent and Henderson, Peter and Islam, Riashat and Bellemare, Marc G and Pineau, Joelle and others},
  journal={Foundations and Trends{\textregistered} in Machine Learning},
  volume={11},
  number={3-4},
  pages={219--354},
  year={2018},
  publisher={Now Publishers, Inc.}
}

@article{towers2024gymnasium,
  title={Gymnasium: A standard interface for reinforcement learning environments},
  author={Towers, Mark and Kwiatkowski, Ariel and Terry, Jordan and Balis, John U and De Cola, Gianluca and Deleu, Tristan and Goul{\~a}o, Manuel and Kallinteris, Andreas and Krimmel, Markus and KG, Arjun and others},
  journal={arXiv preprint arXiv:2407.17032},
  year={2024}
}

@article{sugiyama2008traffic,
  title={Traffic jams without bottlenecks—experimental evidence for the physical mechanism of the formation of a jam},
  author={Sugiyama, Yuki and Fukui, Minoru and Kikuchi, Macoto and Hasebe, Katsuya and Nakayama, Akihiro and Nishinari, Katsuhiro and Tadaki, Shin-ichi and Yukawa, Satoshi},
  journal={New journal of physics},
  volume={10},
  number={3},
  pages={033001},
  year={2008},
  publisher={IOP Publishing}
}

@article{tadaki2013phase,
  title={Phase transition in traffic jam experiment on a circuit},
  author={Tadaki, Shin-ichi and Kikuchi, Macoto and Fukui, Minoru and Nakayama, Akihiro and Nishinari, Katsuhiro and Shibata, Akihiro and Sugiyama, Yuki and Yosida, Taturu and Yukawa, Satoshi},
  journal={New Journal of Physics},
  volume={15},
  number={10},
  pages={103034},
  year={2013},
  publisher={IOP publishing}
}

@inproceedings{kapsalis2025cooperative,
  title={Cooperative Handling of Cut-Ins and Emergency Braking for Connected Automated Vehicles with Heterogeneous Powertrains},
  author={Kapsalis, Dimitris and Spring, John and Lu, Xiao-Yun},
  booktitle={2025 IEEE 64th Conference on Decision and Control (CDC)},
  pages={2152--2159},
  year={2025},
  organization={IEEE}
}

@article{Chandler1958,
 author = {Chandler,R.E. and Herman,R. and Montroll,E.W.},
 journal = {Op. Res.},
 number = {2},
 pages = {165--184},
 title = {Traffic Dynamics: Studies in Car Following},
 volume = {6},
 year = {1958}
}

@article{arnaout2011towards,
  title={Towards reducing traffic congestion using cooperative adaptive cruise control on a freeway with a ramp},
  author={Arnaout, Georges and Bowling, Shannon},
  journal={Journal of Industrial Engineering and Management (JIEM)},
  volume={4},
  number={4},
  pages={699--717},
  year={2011},
  publisher={Barcelona: OmniaScience}
}

@article{brunner2022comparing,
  title={Comparing the observable response times of {ACC} and {CACC} systems},
  author={Brunner, Johannes S and Makridis, Michail A and Kouvelas, Anastasios},
  journal={IEEE Transactions on Intelligent Transportation Systems},
  volume={23},
  number={10},
  pages={19299--19308},
  year={2022},
  publisher={IEEE}
}

@article{jiang2014traffic,
  title={Traffic experiment reveals the nature of car-following},
  author={Jiang, Rui and Hu, Mao-Bin and Zhang, HM and Gao, Zi-You and Jia, Bin and Wu, Qing-Song and Wang, Bing and Yang, Ming},
  journal={PloS one},
  volume={9},
  number={4},
  pages={e94351},
  year={2014},
  publisher={Public Library of Science San Francisco, USA}
}

@article{stern2018dissipation,
  title={Dissipation of stop-and-go waves via control of autonomous vehicles: Field experiments},
  author={Stern, Raphael E and Cui, Shumo and Delle Monache, Maria Laura and Bhadani, Rahul and Bunting, Matt and Churchill, Miles and Hamilton, Nathaniel and Pohlmann, Hannah and Wu, Fangyu and Piccoli, Benedetto and others},
  journal={Transportation Research Part C: Emerging Technologies},
  volume={89},
  pages={205--221},
  year={2018},
  publisher={Elsevier}
}

@article{nishi2013theory,
  title={Theory of jam-absorption driving},
  author={Nishi, Ryosuke and Tomoeda, Akiyasu and Shimura, Kenichiro and Nishinari, Katsuhiro},
  journal={Transportation Research Part B: Methodological},
  volume={50},
  pages={116--129},
  year={2013},
  publisher={Elsevier}
}

@article{hegyi2005optimal,
  title={Optimal coordination of variable speed limits to suppress shock waves},
  author={Hegyi, Andreas and De Schutter, Bart and Hellendoorn, Johannes},
  journal={IEEE Transactions on intelligent transportation systems},
  volume={6},
  number={1},
  pages={102--112},
  year={2005},
  publisher={IEEE}
}

@inproceedings{hegyi2010dynamic,
  title={Dynamic speed limit control to resolve shock waves on freeways-Field test results of the SPECIALIST algorithm},
  author={Hegyi, Andreas and Hoogendoorn, Serge P},
  booktitle={13th International IEEE Conference on Intelligent Transportation Systems},
  pages={519--524},
  year={2010},
  organization={IEEE}
}

@article{pipes1953operational,
  title={An operational analysis of traffic dynamics},
  author={Pipes, Louis A},
  journal={Journal of applied physics},
  volume={24},
  number={3},
  pages={274--281},
  year={1953},
  publisher={American Institute of Physics}
}

@article{he2025review,
  title={A Review of Stop-and-Go Traffic Wave Suppression Strategies: Variable Speed Limit vs. Jam-Absorption Driving},
  author={He, Zhengbing and Laval, Jorge and Han, Yu and Hegyi, Andreas and Nishi, Ryosuke and Wu, Cathy},
  journal={arXiv preprint arXiv:2504.11372},
  year={2025}
}

@article{wang2021jam,
  title={Jam-absorption driving strategy for improving safety near oscillations in a connected vehicle environment considering consequential jams},
  author={Wang, Shunchao and Li, Zhibin and Cao, Zehong and Jolfaei, Alireza and Cao, Qi},
  journal={IEEE Intelligent Transportation Systems Magazine},
  volume={14},
  number={2},
  pages={41--52},
  year={2021},
  publisher={IEEE}
}

@article{gunter2020commercially,
  title={Are commercially implemented adaptive cruise control systems string stable?},
  author={Gunter, George and Gloudemans, Derek and Stern, Raphael E and McQuade, Sean and Bhadani, Rahul and Bunting, Matt and Delle Monache, Maria Laura and Lysecky, Roman and Seibold, Benjamin and Sprinkle, Jonathan and others},
  journal={IEEE Transactions on Intelligent Transportation Systems},
  volume={22},
  number={11},
  pages={6992--7003},
  year={2020},
  publisher={IEEE}
}

@misc{orosz2010traffic,
  title={Traffic jams: dynamics and control},
  author={Orosz, G{\'a}bor and Wilson, R.E. and St{\'e}p{\'a}n, G{\'a}bor},
  journal={Philosophical Transactions of the Royal Society A: Mathematical, Physical and Engineering Sciences},
  volume={368},
  number={1928},
  pages={4455--4479},
  year={2010},
  publisher={The Royal Society Publishing}
}

@article{gasser2004bifurcation,
  title={Bifurcation analysis of a class of ‘car following’ traffic models},
  author={Gasser, Ingenuin and Sirito, Gabriele and Werner, Bodo},
  journal={Physica D: Nonlinear Phenomena},
  volume={197},
  number={3-4},
  pages={222--241},
  year={2004},
  publisher={Elsevier}
}

@article{orosz2004global,
  title={Global bifurcation investigation of an optimal velocity traffic model with driver reaction time},
  author={Orosz, G{\'a}bor and Wilson, R.E. and Krauskopf, Bernd},
  journal={Physical Review E},
  volume={70},
  number={2},
  pages={026207},
  year={2004},
  publisher={APS}
}

@article{treiber2001microsimulations,
title = {Microsimulations of Freeway Traffic Including Control Measures},
author = {M. Treiber and D. Helbing},
pages = {478},
volume = {49},
number = {11},
journal = {at - Automatisierungstechnik},
doi = {doi:10.1524/auto.2001.49.11.478},
year = {2001},
}

@article{davis2004effect,
  title={Effect of adaptive cruise control systems on traffic flow},
  author={Davis, LC},
  journal={Physical Review E},
  volume={69},
  number={6},
  pages={066110},
  year={2004},
  publisher={APS}
}

@techreport{bose2001analysis,
  title={Analysis of traffic flow with mixed manual and intelligent cruise control vehicles: Theory and experiments},
  author={Bose, Arnab and Ioannou, Petros},
  year={2001},
  institution={California Partners for Advanced Transit and Highways}
}

@inproceedings{makridis2018estimating,
  title={Estimating reaction time in adaptive cruise control system},
  author={Makridis, Michail and Mattas, Konstantinos and Borio, Daniele and Giuliani, Raimondo and Ciuffo, Biagio},
  booktitle={2018 IEEE intelligent vehicles symposium (IV)},
  pages={1312--1317},
  year={2018},
  organization={IEEE}
}

@article{makridis2020empirical,
  title={Empirical study on the properties of adaptive cruise control systems and their impact on traffic flow and string stability},
  author={Makridis, Michail and Mattas, Konstantinos and Ciuffo, Biagio and Re, Fabrizio and Kriston, Akos and Minarini, Fabrizio and Rognelund, Greger},
  journal={Transportation research record},
  volume={2674},
  number={4},
  pages={471--484},
  year={2020},
  publisher={SAGE Publications Sage CA: Los Angeles, CA}
}

@book{treiber2012traffic,
  title={Traffic flow dynamics: data, models and simulation},
  author={Treiber, Martin and Kesting, Arne},
  year={2012},
  publisher={Springer Science \& Business Media}
}

@inproceedings{de2004design,
  title={Design and test of a cooperative adaptive cruise control system},
  author={De Bruin, Dik and Kroon, Joris and Van Klaveren, Richard and Nelisse, Martin},
  booktitle={IEEE Intelligent Vehicles Symposium, 2004},
  pages={392--396},
  year={2004},
  organization={IEEE}
}

@inproceedings{bu2010design,
  title={Design and field testing of a cooperative adaptive cruise control system},
  author={Bu, Fanping and Tan, Han-Shue and Huang, Jihua},
  booktitle={Proceedings of the 2010 American Control Conference},
  pages={4616--4621},
  year={2010},
  organization={IEEE}
}

@article{milanes2013cooperative,
  title={Cooperative adaptive cruise control in real traffic situations},
  author={Milan{\'e}s, Vicente and Shladover, Steven E and Spring, John and Nowakowski, Christopher and Kawazoe, Hiroshi and Nakamura, Masahide},
  journal={IEEE Transactions on intelligent transportation systems},
  volume={15},
  number={1},
  pages={296--305},
  year={2013},
  publisher={IEEE}
}

@article{treiber2000congested,
  title={Congested traffic states in empirical observations and microscopic simulations},
  author={Treiber, Martin and Hennecke, Ansgar and Helbing, Dirk},
  journal={Physical Review E},
  volume={62},
  number={2},
  pages={1805},
  year={2000},
  publisher={APS}
}

@misc{SimBufferingStrategy,
  title = {Online simulation module: Stop-and-Go Mitigation via Cooperative Buffering},
  howpublished = {\url{https://www.vzu.uni-wuppertal.de/fileadmin/site/vzu/Stop-and-Go_Mitigation_via_Cooperative_Buffering.html?speed=0.4}},
  note = {Accessed: 2025-11-03}
}

@misc{SimMultiBufferingStrategy,
  title = {Online simulation module: Stop-and-Go Mitigation via Cooperative Multi-Buffering},
  howpublished = {\url{https://www.vzu.uni-wuppertal.de/fileadmin/site/vzu/Stop-and-Go_Mitigation_via_Cooperative_Multi-Buffering.html?speed=0.4}},
  note = {Accessed: 2025-11-03}
}

@misc{SGW_Blog,
  author={Beaty,W.},
  title = {Traffic “experiments” and a cure for waves \& jams,},
  howpublished = {\url{ http://trafficwaves.org/
trafexp.html}},
  note = {Accessed: 2025-10-08}
}

@standard{ISO15622_2018,
  title        = {Intelligent transport systems - {Adaptive Cruise Control} systems — {P}erformance requirements and test procedures, 2nd {E}dition},
  number       = {ISO 15622},
  edition      = {2},
  year         = {2018},
  month        = apr,
  organization = {International Organization for Standardization (ISO)},
  address      = {Geneva, Switzerland},
}

@article{schulman2017proximal,
  title={Proximal policy optimization algorithms},
  author={Schulman, John and Wolski, Filip and Dhariwal, Prafulla and Radford, Alec and Klimov, Oleg},
  journal={arXiv preprint arXiv:1707.06347},
  year={2017}
}

@misc{DRL_code,
  title = {Code of the {DRL} experiment in github},
  author={Korbmacher, Raphael},
  howpublished = {\url{https://github.com/RaphaelVZU/DRL_Vehicle_SGW_Experiement}},
  year = {2025},
  note = {Accessed: 2025-11-12}
}

@article{hayat2025traffic,
  title={Traffic smoothing using explicit local controllers: Experimental evidence for dissipating stop-and-go waves with a single automated vehicle in dense traffic},
  author={Hayat, Amaury and Alanqary, Arwa and Bhadani, Rahul and Denaro, Christopher and Weightman, Ryan J and Xiang, Shengquan and Lee, Jonathan W and Bunting, Matthew and Gollakota, Anish and Nice, Matthew W and others},
  journal={IEEE Control Systems},
  volume={45},
  number={1},
  year={2025},
  publisher={University of California, Berkeley, CA (United States); Vanderbilt~…}
}

@phdthesis{krauss1998microscopic,
  author = {Krau{\ss}, Stefan},
  title  = {Microscopic Modeling of Traffic Flow: Investigation of Collision Free Vehicle Dynamics},
  school = {University of Cologne},
  year   = {1998},
  note = {{DLR}-Forschungsbericht 98-08}
}

\end{document}